\documentstyle[12pt,epsfig]{article}
\newlength{\dinwidth}
\newlength{\dinmargin}
\newcommand{\ba}{\begin{array}}
\newcommand{\ea}{\end{array}}
\newcommand{\beq}{\begin{equation}}
\newcommand{\eeq}{\end{equation}}
\newcommand{\bea}{\begin{eqnarray}}
\newcommand{\eea}{\end{eqnarray}}

\def\ep{\varepsilon}

\def\S{{\bf S}}

\def\bce{\begin{center}}
\def\ece{\end{center}}

\def\nonu{\nonumber}

\def\pa{\partial}
\def\al{\alpha}
\def\be{\beta}
\def\ga{\gamma}

\def\de{\delta}
\def\De{\Delta}
\def\ep{\epsilon}

\def\la{\lambda}
\def\La{\Lambda}

\def\si{\sigma}

\def\S{{\bf S}}

\begin{document}
\thispagestyle{empty}
\addtocounter{page}{-1}
\begin{flushright}
{\tt hep-th/0011121}\\
\end{flushright}
\vspace*{1.3cm}
\centerline{\Large \bf Supersymmetric Domain Wall and RG Flow }
\vskip0.3cm
\centerline{\Large \bf from 4-Dimensional Gauged  ${\cal N}=8$ Supergravity}
\vspace*{1.5cm} 
\centerline{\bf Changhyun Ahn {\rm and} Kyungsung Woo }
\vspace*{1.0cm}
\centerline{\it Department of Physics, 
Kyungpook National University, Taegu 702-701 Korea}
\vskip0.3cm
\vspace*{0.8cm}
\centerline{\tt ahn@knu.ac.kr, \qquad a0418008@rose0.knu.ac.kr}
\vskip2cm
\centerline{\bf abstract}
\vspace*{0.5cm}

By studying various, known extrema of 1) $SU(3)$ sectors, 2) 
$SO(5)$ sectors and 3) $ SO(3) \times
SO(3)$ sectors of  gauged ${\cal N} =8$ supergravity in four-dimensions,
one finds that the deformation of seven sphere
$\S^7$ gives rise to non-trivial 
renormalization group(RG) flow in three-dimensional boundary conformal
field theory 
from UV fixed point to IR fixed point. 
For $SU(3)$ sectors, this leads to four-parameter 
subspace of the supergravity scalar-gravity
action
and we identify one of the eigenvalues
of $A_1$ tensor of the theory with a superpotential of scalar potential 
that governs RG flows on this subspace.
We analyze some of the structure of the superpotential and
discuss first-order BPS domain-wall solutions, using some algebraic
relations between  superpotential and derivatives of it with respect to 
fields, that determine 
a (super)symmetric kink 
solution in four-dimensional ${\cal N} =8$ supergravity, 
which generalizes all the previous 
considerations. 
The BPS domain-wall solutions are equivalent to vanishing of variation
of spin $1/2, 3/2$ fields in the supersymmetry preserving
bosonic background of gauged ${\cal N}=8$ supergravity.
For $SO(5)$ sectors, there exist only nontrivial nonsupersymmetric
critical points that are unstable and included in $SU(3)$ sectors. 
For $SO(3) \times SO(3)$ sectors, we construct the scalar potential(never
been written) explicitly
and study explicit construction of  first-order domain-wall solutions.


\baselineskip=18pt
\newpage

\section{Introduction}
\setcounter{equation}{0}

Few examples are known for three-dimensional interacting conformal field 
theories, mainly due to strong coupling dynamics in the infrared(IR) limit. 
In the previous papers \cite{ar,ar1},  
three-dimensional (super)conformal field theories were classified by utilizing 
the AdS/CFT correspondence \cite{maldacena,witten,gkp} and earlier, exhaustive 
study of the Kaluza-Klein supergravity \cite{duff}.

In contrast to the Freund-Rubin compactifications, the symmetry 
of the vacuum of Englert type compactification is no longer given by
the isometry group of seven-dimensional internal space
 but rather by the group which leaves invariant
{ \it both} the metric and four-form magnetic field strength. By 
generalizing compactification vacuum ansatz to the nonlinear level,
solutions of the eleven-dimensional supergravity were obtained 
directly from the scalar and pseudo-scalar expectation values at
various critical points of the ${\cal N}=8$ supergravity potential \cite{dnw}. 
They reproduced all known Kaluza-Klein solutions of 
the eleven-dimensional supergravity: round $\S^7$ \cite{dp}, 
$SO(7)^-$-invariant, {\sl parallelized} $\S^7$ \cite{englert},
$SO(7)^{+}$-invariant vacuum \cite{dn}, $SU(4)^{-}$-invariant vacuum 
\cite{pw}, and a new one with $G_2$ invariance. 
Among them, round $\S^7$- and $G_2$-invariant vacua are
stable, while $SO(7)^{\pm}$-invariant ones are known to be unstable 
\cite{dn1}. In \cite{ar1}, via AdS/CFT correspondence, deformation of 
$\S^7$ was interpreted as renormalization group flow from ${\cal N}=8$,
$SO(8)$ invariant ultraviolet(UV) fixed point to ${\cal N}=1$, $G_2$ invariant
IR fixed point by analyzing de Wit-Nicolai potential.

Since embedding or consistent truncation of gauged supergravity is known
for $\S^7$ compactification of eleven-dimensional supergravity, we also
are interested in domain-wall solution in four-dimensional supergravity.
In \cite{ap},  
a renormalization group flow from ${\cal N}=8$,
$SO(8)$ invariant UV fixed point to ${\cal N}=2$, $SU(3) \times U(1)$ 
invariant IR fixed point was found 
by studying  de Wit-Nicolai potential which is 
invariant under $SU(3) \times U(1)$ group. 
For this interpretation it was crucial 
to know the form of superpotential that was encoded in the structure of
T-tensor of a theory. Moreover, one can proceed this direction for 
${\cal N}=1$ $G_2$
fixed point \cite{ar2}. It turned out that we found first-order BPS equations,
by recognizing some algebraic and essential relation 
between the superpotential and
derivative of it with respect to field, 
whose solutions constitute supersymmetric domain-walls  {\it both} from
direct minimization of energy-functional and from supersymmetry 
transformation rules.

It is natural and illuminating 
to ask whether one can construct the {\it most general} 
superpotential 
for so-far known any critical points in four-dimensional ${\cal N}=8$
gauged supergravity: 1) $SU(3)$-invariant sectors,
2) $SO(5)$-invariant sectors
and 3) $SO(3) \times SO(3)$-invariant sector. In order to find and study
BPS domain-wall solutions by  minimization of energy-functional, one has to
reorganize it into sum of complete squares. Then one should expect
that the scalar potential takes sum of square of physical quantities.    
One important feature of the de Wit-Nicolai $d=4, {\cal N}=8$ supergravity
is that the scalar potential can be written as the difference of two
positive square terms. 
Together with kinetic terms this implies one may construct
energy-functional in terms of sum of complete squares.       

In this paper, we will continue to analyze
various known vacua of four-dimensional ${\cal N}=8$ supergravity, 
developed earlier by Warner \cite{warner} mainly. 
In section 2, after reviewing de Wit-Nicolai scalar potential 
and by explicitly constructing 28-beins  $u^{IJ}_{\;\;\;KL}$
and $v_{IJKL}$ fields, 
which are elements of fundamental 
56-dimensional representation of $E_7$, in terms of scalar, pseudo-scalar
fields, and other two fields parametrizing $SU(2) \times U(1)$ subgroup
of $SU(8)$ of ${\cal N}=8$ supergravity, 
we get $A_1^{\;\;IJ}$ 
and $A_{2,L}^{\;\;\;IJK}$ tensors 
which are new findings and play an important role.  
Then we possess the full Lagrangian which consists of kinetic terms and 
scalar potential terms in terms of a restricted independent 
four-dimensional slice of scalar
manifold.
Moreover one also considers other two invariant sectors.
In section 3,
we identify one of the eigenvalues of $A_1$ tensor with ``superpotential'' of
de Wit-Nicolai scalar potential.
We describe and present some properties of all the critical points in this 
invariant subsector and discuss some of the implications of our results.
We focus on the nontrivial supersymmetric critical points generalizing
the previous results by \cite{ap,ar2}  
 and obtain the BPS domain-wall solutions
from {\it both} direct extremization of energy-density and supersymmetry 
transformation rules. To arrive at this result, in particular,
some algebraic relations of
superpotential that are newly discovered results 
will play an important role because without them one can not
cancel out the cross terms in energy-functional.
We also present an analytic solution for domain-walls in $SO(3) \times SO(3)$
invariant sector when we assume quadratic order in the 
 fluctuation of field.
Finally, in an appendix, there exist some details.  

\section{de Wit-Nicolai Potential}
\setcounter{equation}{0}

de Wit and Nicolai \cite{dn3,dn2} 
constructed a four-dimensional supergravity theory
by gauging the $SO(8)$ subgroup of $E_7$ in the global $E_7$ $\times$
local $SU(8)$ supergravity of Cremmer and Julia \cite{cremmer} by introducing
the appropriate couplings by hand and then constructing
the supersymmetry model by Noether procedure. In common
with Cremmer-Julia theory, this theory contains self-interaction of a single
massless ${\cal N}=8$ supermultiplet of spins 
$(2, 3/2, 1, 1/2, 0^{+}, 0^{-})$ {\sl but} with local $SO(8)$ $\times$ local
$SU(8)$ invariance. 
There is a new parameter, the $SO(8)$ gauge coupling constant $g$ besides
the gravitational constant.  In order to preserve 
the ${\cal N}=8$ supersymmetry, they modified the Cremmer-Julia
Lagrangian and transformation rules by other $g$-dependent terms.
In particular, there was a non-trivial effective potential for the scalars
that is  proportional to the square of the $SO(8)$ gauge coupling.  
It is well known \cite{cj} that the 70 real, physical 
scalars of ${\cal N}=8$ supergravity
parametrize the coset space $E_7/SU(8)$(even though $E_7$ symmetry is broken 
in the gauged theory) since 63 fields$(133-63=70)$ 
may be gauged
away by an $SU(8)$ rotation(maximal compact subgroup of $E_7$) 
and can be described by 
an element ${\cal V}(x)$ of the fundamental 56-dimensional representation
of $E_7$:
\bea
{\cal V}(x)=
\left(
\begin{array}{cc}
u_{ij}^{\;\;IJ}(x) & v_{ijKL}(x)  \\
v^{klIJ}(x) & u^{kl}_{\;\;KL}(x)
\end{array} 
\right), 
\label{56bein}
\eea
where $SU(8)$ index pairs $[ij], \cdots$ and $SO(8)$ index pairs $[IJ], \cdots$
are antisymmetrized and therefore $u_{ij}^{\;\;IJ}$ and  $v_{ijKL}$ fields
are $28 \times 28$ matrices and $x$ is the coordinate on 4-dimensional 
space-time.
Complex conjugation can be done by raising or lowering those
indices, for example,
$(u_{ij}^{\;\;IJ})^{\star}= u^{ij}_{\;\;IJ}$ and so on.
Under local $SU(8)$ and local $SO(8)$, the matrix ${\cal V}(x)$ transforms
as ${\cal V}(x) \rightarrow U(x) {\cal V}(x) O^{-1}(x)$ where 
$U(x) \in SU(8)$ and $O(x) \in SO(8)$ and matrices $U(x)$ and 
$O(x)$ are in the appropriate 56-dimensional representation.
In the gauged supergravity theory, the 28-vectors transform in the
adjoint of $SO(8)$ with resulting {\it non-abelian} 
field strength while, in ungauged
supergravity theory, all the vector fields have abelian gauge symmetries
and these gauge fields are not minimally coupled to the fermions. 
It is known that any ground state leaving the symmetry unbroken is necessarily
$AdS_4$ space with a cosmological constant proportional to $g^2$. 
One cannot 
identify 70 scalars as the Goldstone bosons of $E_7$ breaking to $SU(8)$ 
because $E_7$ is ${\it no}$ longer a symmetry. 

Although the full gauged ${\cal N}=8$ Lagrangian is rather complicated 
\cite{dn2},
the scalar and gravity part of the action  is simple(we are considering
a gravity coupled to scalar field theory since matter fields do not
play a role in domain-wall solutions)
and maybe written as 
\bea
\int d^4 x \sqrt{-g} \left( \frac{1}{2} R 
- \frac{1}{96} \left| A_{\mu}^{\;\;ijkl} \right|^2 - 
V \right),
\label{action}
\eea
where
the scalar kinetic terms are completely antisymmetric and self-dual
in their indices:
\bea
A_{\mu}^{\;\; ijkl} = -2 \sqrt{2} \left( u^{ij}_{\;\;IJ} 
\partial_{\mu} v^{klIJ} -
v^{ijIJ} \partial_{\mu} u^{kl}_{\;\;IJ} \right)
\label{aijkl},
\eea
where $SO(8)$ indices are contracted 
and $\left| A_{\mu}^{\;\;ijkl} \right|^2$ is a product of 
$A_{\mu}^{\;\;ijkl}  $ and its complex conjugation, $A_{\mu,ijkl}  $
as above and $\mu$ is the 4-dimensional space-time index.
Note that the property of self-dual of $A_{\mu}^{\;\; ijkl}$
can not be obtained from directly (\ref{aijkl}) but from 
group theoretical arguments based on $E_7$ Lie algebra. 
Let us define $SU(8)$ so-called T-tensor which is cubic in the 
28-beins   $u_{ij}^{\;\;IJ}$ and  $v_{ijKL}$ fields, 
manifestly antisymmetric in the indices
$[ij]$ and $SU(8)$ covariant: 
\bea
T_l^{\;kij} & = & 
\left(u^{ij}_{\;\;IJ} +v^{ijIJ} \right) \left( u_{lm}^{\;\;\;JK} 
u^{km}_{\;\;\;KI}-v_{lmJK} v^{kmKL} \right).
\label{ttensor}
\eea
It is not
$E_7$- but only $SO(8)$-invariant since the capital indices 
are contracted in (\ref{ttensor}). 
This comes naturally from introducing
a local gauge coupling in the theory. Furthermore, other tensors coming from
T-tensor play an important role in this paper and scalar structure is
encoded in two $SU(8)$ tensors. 
These appear in the $g$-dependent interaction terms in addition to
the original Lagrangian. 
That is, 
$A_1^{\;\;ij}$ tensor is symmetric in $(ij)$
and $A_{2l}^{\;\;\;ijk}$ tensor is antisymmetric in $[ijk]$: 
\bea
A_1^{\;\;ij} & =& 
-\frac{4}{21} T_{m}^{\;\;ijm}, \;\;\; A_{2l}^{\;\;\;ijk}=-\frac{4}{3}
T_{l}^{\;[ijk]},
\label{a1a2}
\eea
obtained by 
making use of some identities in T-tensor and projecting out the appropriate
irreducible components.

Then de Wit-Nicolai effective nontrivial
potential, which is invariant under the gauged subalgebra, $SO(8)$
of $E_7$, arising from $SO(8)$ gauging can be written as the difference
of two positive definite terms:
\bea
V= -g^2 \left( \frac{3}{4} \left| A_1^{\;ij} \right|^2-\frac{1}{24} \left|
A^{\;\;i}_{2\;\;jkl}\right|^2 \right), 
\label{V}
\eea
where
$g$ is a $SO(8)$ gauge coupling constant and it is understood that
the squares of absolute values of $A_1^{\;ij}, A_{2 \;\;jkl}^{\;\;i}$ 
are nothing but products of
those and their complex conjugations on 28-beins  
$u_{ij}^{\;\;IJ}$ and  $v_{ijKL}$ fields.
The 56-bein ${\cal V}(x)$ can be brought into the following form in the
$SU(8)$ unitary gauge by the 
gauge freedom of $SU(8)$ rotation
\bea
{\cal V}(x)=
\mbox{exp} \left(
\begin{array}{cc}
0 &  \phi_{ijkl}(x)  \\
 \phi^{ijkl}(x) & 0 
\end{array} \right),
\label{calV}
\eea  
where $\phi^{ijkl}$ is a complex self-dual tensor describing the 35 
scalars $\bf 35_{v}$(the real part of $\phi^{ijkl}$)
and 35 pseudo-scalar fields $\bf 35_{c}$(the imaginary part of $\phi^{ijkl}$)
  of ${\cal N}=8$ supergravity.
After gauge fixing, one does not distinguish between $SO(8)$ and $SU(8)$
indices $[IJ]$ and $[ij]$, and they are on equal footing. 
The full supersymmetric solution where ${\it both}$ $\bf 35_{v} $ 
scalars and $\bf 35_{c} $ pseudo-scalars
vanish yields $SO(8)$ vacuum state with ${\cal N}=8$ supersymmetry(Note 
that $SU(8)$ is {\it not} a symmetry of the vacuum). In this case,
70 scalars(and pseudo-scalars) are tachyonic.

\subsection{$SU(3)$ Sectors of Gauged ${\cal N}=8$ Supergravity }

We will start with gauged ${\cal N}=8$ supergravity in four-dimensions.
The scalar potential
is a function of 70 scalars and this number is too large to be managed
practically and one should reduce the problem by looking at 
all critical points that reduce the gauge/R-symmetry to a group containing
a particular $SU(3)$ subgroup of $SO(8)$. 
For one possible embedding of $SU(3)$ corresponding to
the decomposition of three basic representations of $SO(8)$
into $SU(3)$ representations
$\bf 8_v, 8_s, 8_c \rightarrow 3 +\overline{3} + 1 + 1$, all of the 
35-dimensional representations of $SO(8)$ decompose into
$\bf 8 + 6 + \overline{6} + 3 +3 + \overline{3} +
\overline{3} + 1 + 1 + 1$. Then the set of 70 scalars in ${\cal N}=8$
supergravity containes 6 singlets of $SU(3)$(three singlets for $\bf 35_v$
and three singlets for $\bf 35_c$). 
For the other embeddings of $SU(3)$ 
$\bf 8_v, 8_s, 8_c \rightarrow 8 $, 
all of the 
35-dimensional representations of $SO(8)$ decompose into
$\bf 27 + 8 $ implying that there are no $SU(3)$ singlets in the 
scalar sectors.
It is known \cite{warner} 
that $SU(3)$ singlet space with a breaking of the $SO(8)$ gauge group
into a group which contains $SU(3)$
may be written in terms of  two real parameters $\la$ and $\la'$:
\bea 
\phi_{ijkl} = S(\la G^{+}_{1} + \la' G^{+}_{2}), 
\nonu
\eea 
where the action $S$ is $SU(2) \times U(1)$ subgroup of $SU(8)$
on its 70-dimensional representation in the space of self-dual 
complex four-forms:
\bea 
S=\mbox{diag}(w,w,w,w,w,w,w^{-3}P),
\nonu
\eea 
where
$w=e^{i\alpha/4}$ is a pure phase, and $P$ is a general $SU(2)$
matrix: 
\bea 
P & = & \left(\begin{array}{cc} \mbox{cos}\;\theta &
- \mbox{sin}\;\theta \nonu \\ \mbox{sin}\;\theta &
\mbox{cos}\;\theta
\end{array} \right) \left(
\begin{array}{cc} e^{i\phi} & 0 \nonu \\ 0 & e^{-i\phi}
\end{array} \right) \left(
\begin{array}{cc} \mbox{cos}\;\psi & - \mbox{sin}\;\psi \nonu \\
\mbox{sin}\;\psi & \mbox{cos}\psi
\end{array} \right). 
\nonu
\eea
In the notation
of \cite{warner1}, $G^{+}_{1}=X_{1}^{+}+X_{2}^{+}+X_{3}^{+}$ and
$G^{+}_{2}=X_{4}^{+}+X_{5}^{+}+X_{6}^{+}+X_{7}^{+}$  where 
the self-dual and anti-self-dual four-forms are given by 
\bea
& & X_1^{\pm} = \frac{1}{2} (\de^{1234}_{ijkl} \pm
\de^{5678}_{ijkl}), 
X_2^{\pm} =\frac{1}{2}
 (\de^{1256}_{ijkl} \pm \de^{3478}_{ijkl}), 
X_3^{\pm}= \frac{1}{2} (\de^{1278}_{ijkl} \pm \de^{3456}_{ijkl}), \nonu \\
& & X_4^{\pm} = -\frac{1}{2}(\de^{1357}_{ijkl}
\pm \de^{2468}_{ijkl}), 
X_5^{\pm} = \frac{1}{2} (\de^{1368}_{ijkl} \pm \de^{2457}_{ijkl}),
X_6^{\pm} = \frac{1}{2} (\de^{1458}_{ijkl} \pm \de^{2367}_{ijkl}), 
\nonu \\
& & X_7^{\pm} = \frac{1}{2}
 (\de^{1467}_{ijkl} \pm \de^{2358}_{ijkl}). 
\label{xplus}
\eea

Then the parametrization of \cite{warner} for the $SU(3)$-singlet space 
that is invariant subspace under a particular $SU(3)$ subgroup
of $SO(8)$
becomes
\begin{eqnarray}
\phi_{ijkl} & = &  \; \la \; \mbox{cos}\alpha\; Y^{1\;+} +
\la \; \mbox{sin}\alpha\; Y^{1\;-} + \la' \; \mbox{cos} \phi \;
\; Y^{2\;+} 
 + \la' \; \mbox{sin} \phi \; 
Y^{2\;-} \; ,
\label{phiijkl}
\end{eqnarray}
where
\begin{eqnarray}
  Y^{1\;\pm}_{ijkl} &=& \varepsilon_{\pm} \left[ \; (\de^{1234}_{ijkl} \pm 
\de^{5678}_{ijkl})+
 (\de^{1256}_{ijkl} \pm \de^{3478}_{ijkl})+(\de^{3456}_{ijkl}
 \pm \de^{1278}_{ijkl}) \; \right],
\nonu \\
       Y^{2\;\pm}_{ijkl} &=& \varepsilon_{\pm} \left[ -(\de^{1357}_{ijkl}
\pm \de^{2468}_{ijkl})+(\de^{2457}_{ijkl} \pm
\de^{1368}_{ijkl})+(\de^{2367}_{ijkl} \pm \de^{1458}_{ijkl}) +
 (\de^{1467}_{ijkl} \pm \de^{2358}_{ijkl}) \right], 
\label{Y}
\end{eqnarray} 
where $\varepsilon_{+}=1$ and $\varepsilon_{-}=i$ and $+$ gives the scalars
and $-$ the pseudo-scalars.
Therefore 56-beins ${\cal V}(x)$ can be written as $ 56\times 56$ matrix whose
elements are some functions of scalar, pseudo-scalars, $\al$ and $\phi$
out of seventy fields
by exponentiating
the vacuum expectation value $\phi_{ijkl}$ through (\ref{calV}). 
On the other hand, 28-beins $u^{IJ}_{\;\;\;KL}$
and $v_{IJKL}$ are 
elements of this ${\cal V}(x)$ according to (\ref{56bein}).
One can construct 28-beins $u^{IJ}_{\;\;\;KL}$ and $v_{IJKL}$ 
in terms of these fields
explicitly and they are given in the appendix (\ref{uvsu3}). 
Now it is ready to get the complete expression for $A_1^{IJ}$ and 
$A_{2, L}^{\;\;\;\;\;IJK}$ tensors in terms of $\la, \la', \al$ and 
$\phi$ using
(\ref{ttensor}) and (\ref{a1a2}). 
This will be our first new findings and main ingradients that are necessary
to proceed further.

It turns out from (\ref{a1a2}) that 
$A_1^{\;IJ}$ tensor has three distinct complex
eigenvalues, $z_1$, $z_2$ and $z_3$ 
with degeneracies 6, 1, and 1 respectively and has the following form
\begin{equation}
A_1^{\;IJ} =\mbox{diag} \left(z_1, z_1, z_1, z_1, z_1,
z_1, z_2, z_3 \right), \label{a1}
\nonu
\end{equation}
where the eigenvalues are functions of $\la, \la', \al$ and $\phi$:
\begin{eqnarray}
z_1 & = & e^{-2i(\alpha + \phi)} \left[e^{3i\alpha}p^2qr^2t^2 +
e^{i(3\alpha+4\phi)}p^2qr^2t^2 +  e^{2i (\alpha +
\phi)}p \left(4q^2r^2t^2+p^2 \left(r^4 + t^4 \right) \right) \right.  \nonu \\
 & &  \left. + pq^2r^2t^2  +
e^{4i\phi}pq^2r^2t^2 + e^{i(\alpha + 2\phi)}q \left(4p^2r^2t^2 + q^2
\left(r^4
+ t^4 \right) \right) \right], \nonu \\ 
z_2 & = & e^{-4i\phi} \left(p +
e^{i\alpha}q \right) \left(e^{4i\phi}p^2r^4 -e^{i(\alpha + 4\phi)}pqr^4 +
e^{2i(\alpha + 2\phi)}q^2r^4 + 6e^{i(\alpha + 2\phi)}pqr^2t^2
\right. \nonu \\ 
& & \left. + p^2t^4 - e^{i\alpha}pqt^4 + e^{2i\alpha}q^2t^4 \right), \nonu \\
z_3 & = & 6e^{i(\alpha + 2\phi)}p^2qr^2t^2
 + 6e^{2i(\alpha + \phi)}pq^2r^2t^2 + p^3 \left(r^4 + e^{4i\phi}t^4 \right)  +
e^{3i\alpha}q^3 \left(r^4 + e^{4i\phi}t^4 \right),
\nonu
\end{eqnarray}
and we denote hyperbolic functions of $\la$ and $\la'$ by the following 
quantities which will be used throughout this paper
\bea 
& & p \equiv \cosh \left(\frac{\la}{2\sqrt{2}}\right), \;\; q \equiv
\sinh\left(\frac{\la}{2\sqrt{2}}\right), 
\;\; r \equiv \cosh\left(\frac{\la'}{2\sqrt{2}}\right), \;\; t
\equiv \sinh\left(\frac{\la'}{2\sqrt{2}}\right). 
\label{pqrt}
\eea 
One of the eigenvalues of
$A_1^{\;\;IJ}$ tensor, $z_3$,  will provide a ``superpotential'' of scalar
potential $V$ and
be crucial for analysis of domain-wall solutions later.
First, the BPS domain-wall solutions are nothing but
the gradient flow equations of this superpotential defined on
a four-dimensional slice of the full scalar manifold. Second, 
the modified $g$-dependent supersymmetry transformation rule of gravitinos 
obtained by gauging $SO(8)$ group 
contains the superpotential and it is very important to
have this form of superpotential when we consider its properties under the
supersymmetric bosonic background.
   
Similarly, $A_{2, L}^{\;\;\;\;\;IJK}$ 
tensor can be obtained from the triple product of
$u^{IJ}_{\;\;\;KL}$ and $v_{IJKL}$ fields, that is, from (\ref{a1a2}). 
It turns out that they are written as eight-kinds of 
fields
$y_i$ where $i=1, \cdots, 8$ and are given in the appendix (\ref{su3y})
where we stressed the fact that some of these are related to 
the derivatives of eigenvalues of $A_1^{IJ}$ tensor with respect to
$\la$ and $\la'$.
As already mentioned before, the scalalr potential consists of
$\left| A_1^{\;ij} \right|^2$ and $\left|
A^{\;\;i}_{2\;\;jkl}\right|^2$. Since the former is made of squares of 
superpotential plus other terms and the latter is made of 
squares of derivatives of superpotential with respect to $\la$
and $\la'$ plus other terms, we will see that both other terms from $A_1$
and $A_2$ tensors are exactly cancelled out and lead to the sum of
square of superpotential and
square of derivatives of superpotential.
Finally, 
the scalar potential (\ref{V})
can be written, by combining all the components of
$A_1^{IJ}, A_{2, L}^{\;\;\;\;\;IJK} $ tensors,  as
\bea 
&& V(\la, \la', \al, \phi)  
= -g^2 \left[ \; \frac{3}{4} \times \left( 6|z_1|^2 + |z_2|^2 + |z_3|^2
\right) \right. \nonu \\
&&  \left.  - \frac{1}{24} \times 6 \left( 12|y_1|^2 +3|y_2|^2+ 
3|y_3|^2 + 12|y_4|^2 +
12|y_5|^2 + 4|y_6|^2 + 4|y_7|^2 + 6|y_8|^2 \right) \; \right]
\nonu \\ 
&& =  \frac{1}{2} g^2 \left( s'^4 \left[ (x^2+3)c^3 + 4x^2v^3s^3 -
3v(x^2-1)s^3 + 12xv^2cs^2 - 6(x-1)cs^2 + 6(x+1)c^2sv \right] \right.  \nonu \\ 
&& \left. + 2s'^2 \left[ 
2(c^3+v^3s^3) + 3(x+1)vs^3 + 6xv^2cs^2 - 3(x-1)cs^2 -
6c \right]- 12c \right), 
\label{potential} 
\eea 
which is exactly the same form obtained by \cite{warner} using $SU(8)$
coordinate as an alternative approach for which 
one has to know about kinetic terms
explicitly as well as scalar potential terms 
in order to understand the supergravity domain-wall solutions
and we introduce
the following quantities for simplicity
\bea & & c \equiv
\cosh\left(\frac{\la}{\sqrt{2}}\right), \; s \equiv \sinh\left(\frac{
\la}{\sqrt{2}}\right), 
\; c' \equiv
\cosh\left(\frac{\la'}{\sqrt{2}}\right), \; s' \equiv 
\sinh\left(\frac{\la'}{\sqrt{2}}\right) \nonu
\\ & & v \equiv \cos\alpha ,\;\;\;\;\;\;\;\;\;\;\; x \equiv 
\cos2\phi.   
\label{cs} 
\eea
Although one gets the explicit form of scalar potential 
by exploiting the method given by \cite{warner}, another task 
is to find out kinetic terms. This is one of the reasons why we took different 
route.
The scalar potential does not depend on $\theta$ and $\psi$ of $SU(2)$
matrix reflecting
$SO(8)$ invariance of the potential and a larger invariance of the 
$SU(3)$-singlet sector, respectively. 
The potential contains as special case the examples previously studied
in the literature.
One can easily see that by putting
$\la=\la'$ and $\al =\phi$, (\ref{potential}) will reduce to the one 
studied in \cite{ar2,ar1} while
by putting $\al=0$ and $\phi=\pi/2$, one gets the one considered in 
\cite{ap}.

\subsection{$SO(5)$ Sectors of Gauged ${\cal N}=8$ Supergravity }

The non-maximally symmetric 
example of the Freund-Rubin compactification to a product
of $AdS_4$ space-time and an aritrary compact seven-dimensional Einstein
manifold is provided by squashed seven sphere $\S^7$. 
The effective four-dimensional 
theory has $SO(5) \times SO(3)$ gauge symmetry and ${\cal N}=1$ or
${\cal N}=0$ depending on the orientation of the $\S^7$ \cite{awada}.
The original motivation for studying all the critical points of
gauged ${\cal N}=8$ supergravity having $SO(5)$ symmetry at least was to find  
some connections between Freund-Rubin type solution and de Wit-Nicolai theory.
One must characterize the action of $SO(5) \subset 
SO(8)$ on the physical fields.
There are three ways of embedding $SO(5)$ in $SO(8)$. In the symmetric gauge,
the seventy scalars of ${\cal N}=8$ supergravity maybe described as 
elements of self-dual four forms and anti-self-dual four forms. 
The 35 scalars break into $\bf 35_v \rightarrow 10 + 10 + 10 +5$ of $SO(5)$
while 35 pseudo-scalars $\bf 35_c \rightarrow 10 + 10 + 10 + 5$.
Therefore $SO(5)^{v}$ embedding case gives no $SO(5)$-singlets among
either true scalars or pseudo-scalars where $\bf 8_v \rightarrow 5 + 
1+ 1 + 1$. Then $SO(5)^{v}$ embedding gives only trivial vacuum where all
seventy scalars vanish with unbroken $SO(8)$ symmetry. We restrict to ourselves for
other two embeddings 1) $SO(5)^{+}$ embedding case where $\bf 8_s \rightarrow
 5 + 1 + 1 +1$ and 2) $SO(5)^{-}$ embedding case
where $\bf 8_c \rightarrow 5 + 1 + 1 + 1$.

$\bullet$ $SO(5)^{+}$ embedding

This case contains $SO(5)$-singlets among six scalars$(\bf 35_v 
\rightarrow 14 + 5 + 5 + 5 + 1 + 1+ 1+ 1+ 1 +1)$ which can be 
parametrized by the following form after acting $SO(3)$ rotation of 
$SO(8)$ on its 70-dimensional representation in the space:  
\bea 
\phi_{ijkl} = \la \left( X_1^{+}+X_2^{+}+X_3^{+} \right)+
\mu \left(X_1^{+}+X_4^{+}+X_5^{+} \right) + \rho \left(X_1^{+}-
X_6^{+}-X_7^{+} \right), 
\label{phiijklso5}
\eea 
where
$\la, \mu$(not space-time index) and 
$\rho$ are real parameters and self-dual four-forms
$X^{+}_{\al}$'s are given in
(\ref{xplus}). We used the fact that
as a consequence of $SO(8)$ symmetry of the theory, the potential does not
depend on $SO(3)$ rotation parametrized by other three parameters. 
As we have done before, we can describe 28-beins $u^{IJ}_{\;\;\;KL}$ 
and $v_{IJKL}$ in terms of
$\la, \mu$ and $\rho$ and they are given in the appendix (\ref{uvso5+}) with
upper plus sign for each $4 \times 4$ submatrix $u_i^{+}$ and $v_i^{+}$.
It turns out that $A_1^{\;IJ}$ tensor has a single eigenvalue $z_1$ with
multiplicity 8 which will provide a superpotential of scalar potential
 and has the following expression
\bea
A_1^{\;IJ} =\mbox{diag} \left(z_1, z_1, z_1, z_1, z_1,
z_1, z_1, z_1 \right), 
\nonu
\eea
where the eigenvalues are real and some combinations of $u, v$ and $w$ fields,
and  take the form 
\bea
z_1 = -\frac{1}{8\sqrt{uvw}} \left(5 + u^2v^2 + 
\mbox{two cyclic permutations}
 \right),
\label{z1so5+} 
\eea
and we introduce\footnote{Unfortunately we use $u, v$ letters 
here in order to keep the notation as in the literature \cite{romans}. 
We hope these 
are nothing to do with 28-beins $u^{IJ}_{\;\;\;KL},v_{IJKL}$ 
fields we have used before. }
new fields $u,v$ and $w$ as
\bea
u \equiv e^{\la/\sqrt{2}}, \qquad v \equiv e^{\mu/\sqrt{2}},
\qquad w \equiv e^{\rho/\sqrt{2}}.
\label{uvw}
\eea
Also one can construct  $A_{2, L}^{\;\;\;\;\;IJK}$ tensor
which are the combinations of triple product of 28-beins 
$u^{IJ}_{\;\;\;KL}$ and $v_{IJKL}$.
They are written as four-kinds of fields $y_{i,+}$ where $i=1, \cdots, 4$ and 
are given in the appendix (\ref{so5y+}).
Therefore one gets the scalar potential $V(\la, \mu, \rho)$ by summing all the
components of $A_1^{IJ}, A_{2, L}^{\;\;\;\;\;IJK}$ 
tensors and counting the degeneracies correctly:
\bea
V & = & -g^2 \left[ \; \frac{3}{4} \times 8 z_1^2 
 - \frac{1}{24} \times 48 \left(  y_{1,+}^2 +2 y_{2,+}^2+ 2 y_{3,+}^2 + 
2 y_{4,+}^2 
 \right) \; \right] \nonu \\ 
& = & \frac{1}{8} g^2 \left( u^3v^3/w-10
uv/w   -2 uvw^3 + \mbox{two cyclic permutations} -15/uvw \right). 
\nonu
\eea
This is exactly the same form obtained by Romans \cite{romans}.
There exists one nontrivial extremum at $u=v=1/w=5^{1/4}$ which has a 
${\cal N}=0$ nonsupersymmetric  $SO(7)^{+}$ gauge symmetry besides
trivial one which has ${\cal N}=8$ maximal supersymmetric $SO(8)$
gauge symmetry for which $u=v=w=1$.

$\bullet$ $SO(5)^{-}$ embedding

In this case, there exist six $SO(5)$-singlets among the pseudo-scalars
$(\bf 35_c \rightarrow 14 + 5 + 5 + 5 + 1 + 1 + 1 + 1 + 1 + 1)$.
By extra $SU(8)$ element transforming self-dual into anti-self-dual 
four-forms, one can parametrize as follows.
\bea 
\phi_{ijkl} = i \la \left( X_1^{-}+X_2^{-}+X_3^{-} \right)+
i \mu \left(X_1^{-}+X_4^{-}+X_5^{-} \right) + i \rho \left(X_1^{-}-
X_6^{-}-X_7^{-} \right), 
\label{so5-}
\eea 
with (\ref{xplus}).
Similarly,  it turns out  that 
\bea
A_1^{\;IJ} =\mbox{diag} \left(z_1, z_1, z_1, z_1, z_1,
z_1, z_1, z_1 \right), 
\nonu
\eea
where the eigenvalues are complex and are given by
\bea
z_1 = \frac{(1+i)}{16} \frac{1}{(uvw)^{3/2}} \left(
 -i u^2 + u^3 v^3 w +\mbox{two cyclic permutations}+ 5 uvw-5 i u^2v^2w^2  
\right), 
\label{z1so5-} 
\eea
with (\ref{uvw}).
Therefore one gets the scalar potential $V(\la, \mu, \rho)$ by summing over
all the
components of $A_1^{IJ}, A_{2, L}^{\;\;\;\;\;IJK}$ tensors with (\ref{so5y-}):
\bea
V & = & -g^2 \left[ \; \frac{3}{4} \times 8|z_1|^2 
 - \frac{1}{24} \times 48 \left( |y_{1,-}|^2 +2|y_{2,-}|^2+ 
2|y_{3,-}|^2 + 2|y_{4,-}|^2 
 \right) \; \right] \nonu \\ 
& = & \frac{1}{16} g^2 \left[ u^3v^3/w+w/u^3v^3-2(
uvw^3+1/uvw^3) -10(uv/w+w/uv)  \right. \nonu \\
& & \left. + \mbox{two cyclic permutations}
 -15(uvw +1/uvw) \right], 
\nonu
\eea
which was found in \cite{romans} and 
has two nontrivial extrema: one with ${\cal N}=0$ nonsupersymmetric 
$ SO(7)^{-}$ symmetry at $u=v=1/w=
(1+\sqrt{5})/2$ and the other with ${\cal N}=0$ nonsupersymmetric  
$SO(6)^{-}=SU(4)^{-}$ symmetry 
at $u=1/w=\sqrt{2}+1$ and $v=1$.
Since $SO(7)^{\pm}$ and $SO(6)^{-}$ contain $SU(3)$ as a subgroup, 
these critical points
also appeared in the previous subsection for the $SU(3)$ sectors. 

\subsection{$SO(3) \times SO(3)$ Sectors of Gauged ${\cal N}=8$ Supergravity }

de Wit and Nicolai have constructed gauged supergravity theories for
${\cal N}=5$ \cite{dn6} and this form of ${\cal N}=5$ scalar potential was 
obtained by natural truncation of ${\cal N}=8$ scalar 
potential in \cite{warner1}(${\cal N}=6$ scalar potential 
was obtained also). 
Moreover ${\cal N}=4$ gauged $SO(4)$ supergravity was obtained by
truncation of ${\cal N}=8$ gauge $SO(8)$ supergravity \cite{clp}.
It is known \cite{warner1} 
that $SO(3) \times SO(3)$ singlet space with a breaking of the 
$SO(8)$ gauge group
into $SO(3) \times SO(3)$
may be written as\footnote{Our convention for $\la^{\al}$ field is different
from that of Warner: $\la^{\al}_{our} = \la^{\al}_w /\sqrt{2}$. }
\bea 
\phi_{ijkl} = S(\la^{\al} X_{\al}^{+}), \qquad \al=1, \cdots, 7 
\label{so3so3ijkl}
\eea 
where the action $S$ is  $SO(3) \times SO(3)$ subgroup of $SU(8)$
on its 70-dimensional representation in the space of self-dual four-forms:
\bea 
S=\mbox{diag}(1,1,1,P,1,1,1), \qquad  P = 
\frac{1}{\sqrt{2}} \left(\begin{array}{cc} 1 &
i \nonu \\ i &  1
\end{array} \right).
\nonu
\eea 
Self-dual four forms $X^{+}_{\al}$'s are the same as in (\ref{xplus}).
From explicit form of 28-beins $u^{IJ}_{\;\;\;KL}$ and $v_{IJKL}$(
and are 
given in the appendix of original version of hep-th archive), 
those are functions of seven parameters 
$\la^{\al}$ and 
it turns out  that
$A_1^{\;IJ}$ tensor has eight distinct components, $z_i$ where $i=1, \cdots, 
8$ with degeneracies 2 and has the following form
\bea
A_1^{\;IJ} = \left(
\begin{array}{cccccccc}
z_1 &0 & 0& 0&0 &0 &0 & z_2  \nonu \\
0 &z_3 &0 &0 &0 &0 &z_4 & 0 \nonu \\
0 &0 &z_5 &0 &0 &z_6 &0 &0  \nonu \\
0 &0 &0 &z_7 &z_8 &0 &0 &0  \nonu \\
0 &0 &0 &z_8 &z_7 &0 &0 &0  \nonu \\
0 &0 &z_6 & 0& 0&z_5 &0 &0  \nonu \\
0 &z_4 &0 &0 &0 &0 &z_3 &0  \nonu \\
z_2 &0 &0 &0 &0 &0 &0 &z_1  \nonu \\
\end{array} \right), \nonu
\eea
where these components are some functions of $\la^{\al}(\al=1, \cdots, 7)$
as follows:
\bea
z_1 & = & q_1(p_4p_7q_2(p_5p_6q_3+p_3q_5q_6)+q_4(p_2p_3p_6p_7q_5 +
p_2p_5p_7q_3q_6 + p_3p_5p_6q_2q_7 \nonu \\
 & & + q_2q_3q_5q_6q_7)) + (p_i \leftrightarrow q_i), \nonu \\
z_2 & = & -\frac{i}{4} (r_1-r_2-r_4+r_7)(p_6q_3q_5+p_3p_5q_6), \nonu \\
z_3 & = & q_1(q_2q_4(p_3p_5p_6+q_3q_5q_6)q_7+p_4(p_5p_6p_7q_2q_3+
p_3p_7q_2q_5q_6 \nonu \\
& & +p_2p_6q_3q_5q_7+p_2p_3p_5q_6q_7)) +  (p_i \leftrightarrow q_i), \nonu \\
z_4 & = &  -\frac{i}{4} (r_1-r_2+r_4-r_7)(p_6p_3q_5+q_3p_5q_6), \nonu \\
z_5 & = & q_1(q_2q_4(p_3p_5p_6+q_3q_5q_6)q_7+p_2(p_3p_6p_7q_4q_5+
p_5p_7q_3q_4q_6 \nonu \\
& & + p_4p_6q_3q_5q_7)) + (p_i \leftrightarrow q_i), \nonu \\ 
z_6 & = &  -\frac{i}{4} (r_1+r_2-r_4-r_7)(p_6q_3p_5+p_3q_5q_6), \nonu \\
z_7 & = &  \frac{1}{4} (r_1+r_2+r_4+r_7)(p_6p_3p_5+q_3q_5q_6), \nonu \\
z_8 & = &  i (p_4(q_5(p_3p_7q_1q_2q_6+p_1p_3p_6q_2q_7+p_2p_6q_1q_3q_7) \nonu \\
& & + p_5(p_6p_7q_1q_2q_3+p_2p_3q_1q_6q_7+p_1q_2q_3q_6q_7)))
+ (p_i \leftrightarrow q_i), 
\label{z1so3so3}
\eea
and a compact notation can be defined by setting(of course these are nothing 
to do with the one in (\ref{pqrt})):
\bea 
& & p_i \equiv \cosh \left(\frac{\la^i}{2}\right), \;\; q_i \equiv
\sinh\left(\frac{\la^i}{2}\right), 
\;\; r_i \equiv \cosh \la^i, \;\; t_i
\equiv \sinh \la^i, i=1,2, \cdots, 7. 
\label{pqrtso3so3}
\eea 
One of the components of $A_1^{IJ}$ tensor, $z_7$, plays a role of
superpotential of scalar potential.
Similarly, $A_{2, L}^{\;\;\;\;\;IJK}$ 
tensor can be obtained from the triple product of
$u^{IJ}_{\;\;\;KL}$ and $v_{IJKL}$ fields. 
It turns out that they are written as 54-kinds of 
fields
$y_i, y_{i, +}$ and $y_{i, -}$( and are 
given in the appendix of original version of hep-th archive).
Finally, 
the scalar potential as a function of seven $\la^{\al}$'s 
can be written by combining all of the components of
$A_1^{IJ}, A_{2, L}^{\;\;\;\;\;IJK}$ tensors  as
\bea
V  & = &  -g^2 \left[ \; \frac{3}{4} \times 2 \sum_{i=1}^8 |z_i|^2 
 - \frac{1}{24} \times 12 \left( \sum_{i=1}^{4}|y_{i,+}|^2+
 \sum_{i=1}^{4}|y_{i,-}|^2 + \sum_{i=5}^{8}|y_i|^2 +
 2 \sum_{i=9}^{10} |y_i|^2  \right. \right. \nonu \\
& & \left. \left. +  \sum_{i=11}^{18}|y_i|^2 +
 \sum_{i=19}^{34}|y_{i,+}|^2 +  \sum_{i=19}^{34}|y_{i,-}|^2  
\right) \; \right]
\nonu \\
& = & \frac{g^2}{32} ( -2r_6+r_1r_2r_6+r_1r_4r_6 -32 r_2r_4r_6-
32 r_1r_6r_7 +r_2r_6r_7 +r_4r_6r_7 -2r_1r_2r_4r_6r_7 
\nonu \\ 
& & - 16t_1t_2t_3
 +2r_4r_7t_1t_2t_3-16t_1t_4t_5+2r_2r_7t_1t_4t_5+4r_6
t_2t_3t_4t_5 
 + 16r_1r_6r_7t_2t_3t_4t_5 \nonu \\
& &  -16t_2t_4t_6-4r_1r_7t_2t_4t_6-32t_3t_5t_6
+2r_1^2t_3t_5t_6 
- r_1r_2t_3t_5t_6+2r_2^2t_3t_5t_6 \nonu \\
& & -r_1r_4t_3t_5t_6+2r_2r_4t_3t_5t_6+
2r_4^2t_3t_5t_6 
 + 2r_1r_7t_3t_5t_6-r_2r_7t_3t_5t_6-r_4r_7t_3t_5t_6 \nonu \\
& & +24r_1r_2r_4r_7t_3t_5t_6
+ 2r_7^2t_3t_5t_6+2t_1^2t_3t_5t_6+2t_2^2t_3t_5t_6+2t_3t_4^2t_5t_6-
2r_6t_1t_2t_4t_7 \nonu \\
& & - 16t_3t_4t_7 +2r_1r_2t_3t_4t_7-16t_2t_5t_7+2r_1r_4t_2t_5t_7+
4r_6t_1t_3t_5t_7 
 + 16 r_2r_4r_6t_1t_3t_5t_7 \nonu \\
& & -16t_1t_6t_7-4r_2r_4t_1t_6t_7+24t_1t_2t_3t_4
t_5t_6t_7 + 2t_3t_5t_6t_7^2 +r_5(1-2r_6t_1t_2t_3 \nonu \\
& & +16r_4r_6r_7t_1t_2t_3-2t_1t_3t_4t_6
 - r_2(r_4+32r_7-16r_7t_1t_3t_4t_6)+t_1t_2t_4t_7 \nonu \\
& & -2r_6t_3t_4t_7-
2t_2t_3t_6t_7 
 + r_1(-r_7+16r_2r_6t_3t_4t_7 +r_4(-32 +r_2r_7 +16t_2t_3t_6t_7))) \nonu \\
& & + r_3(1-32r_5r_6 +2r_1^2 r_5r_6+2r_2^2r_5r_6+2r_4^2r_5r_6-32r_4r_7-
r_4r_5r_6r_7 + 2r_5r_6r_7^2 \nonu \\
& & +2r_5r_6t_1^2+2r_5r_6t_2^2+2r_5r_6t_4^2-2r_6t_1t_4t_5+4
r_5t_2t_4t_6  -2t_1t_2t_5t_6 \nonu \\
& & +16 r_4r_7t_1t_2t_5t_6+t_1t_2t_4t_7 +24r_5r_6t_1t_2t_4t_7-
2r_6t_2t_5t_7 + 4r_5t_1t_6t_7-2t_4t_5t_6t_7 \nonu \\
& & +2r_5r_6t_7^2+r_1(r_7(-1+2r_5r_6+
16r_5t_2t_4t_6) - r_4r_6(r_5-16t_2t_5t_7) \nonu \\
& & +r_2(-32+r_4r_7+r_5r_6(-1+24r_4r_7)+
16t_4t_5t_6t_7)) \nonu \\
& & + r_2(-r_6r_7(r_5-16t_1t_4t_5)+r_4(-1+2r_5(r_6+8t_1t_6t_7))))), 
\nonu
\eea
with (\ref{pqrtso3so3}).
Although the property of the scalar potential was discussed
in \cite{warner1}, it was never written explicitly.
One can easily calculate 
the derivatives of this scalar potential $V$ with repect to
$\la^i$'s and verify that there exists one nontrivial extremum at 
$\cosh \la^1 =9, \la^{\al} =0$ where $\al=2, \cdots, 7$ besides
trivial one which has ${\cal N}=8$ maximal supersymmetric $SO(8)$ gauge 
symmetry for which all parameters vanish and whose cosmological
constant is $-6g^2$. However 
the cosmological
constant becomes $-14 g^2$ at this extremal surface \cite{warner1} which 
generalizes $SO(5)$ model. The extremal structure of the ${\cal N}=5$
potential is exhibited by the ${\cal N}=8$ potential which breaks 
the $SO(8)$ down to $SO(3) \times SO(3)$. That is, 
the surface of stationary points was obtained by embedding the ${\cal N}=5$
stationary surface in the ${\cal N}=8$ theory.
One finds that, at this nontrivial surface,
the $A_1^{IJ}$ tensor has the following form with degeneracies 6, 2
\bea
A_1^{\;IJ} = \mbox{diag} \left( \sqrt{5}, \sqrt{5}, \sqrt{5},
3, 3, \sqrt{5}, \sqrt{5}, \sqrt{5} \right).
\nonu
\eea
One can easily check that there is no supersymmetry because  the 
eigenvalues$(\sqrt{5} \mbox{or} 3 )$  of
$A_1^{\;\;IJ}$ tensor
are not equal to  $\sqrt{-\Lambda/6 g^2}$ at 
$\La=-14g^2$.   In other words, if there is a supersymmetry, then the 
cosmological constant must be either $-30g^2$ or $-54g^2$. But the 
gravitational field equations require that the $AdS_4$ vacuum on the
extremal surface has $\La =-14 g^2$. Therefore there are no supersymmetries. 

\section{Supersymmetric Domain Wall and RG Flow}

\subsection{$SU(3)$ Sectors }

In this subsection, 
we investigate domain walls \cite{cveticetal}
arising in supergravity
theories with a nontrivial superpotential defined on a restricted 
independent four-dimensional slice of the scalar manifold.
We analyze a particular $SU(3)$ invariant sector of the scalar manifold
of gauged ${\cal N}=8$ supergravity in four-dimensions and
study all the critical points of the potential within this sector.
The critical points give rise to $AdS_4$ vacua and preserve at least 
 $SU(3)$ gauge symmetry in the supergravity(or R-symmetry of the 
dual field theory).
The presence and exact knowledge of the 
supergravity potential implies a completely determined non-trivial operator
algebra in dual field theory.
Using Einstein's equations and energy condition, it will be possible to
show that monotonic function can be found in any kink geometry
with Poincare symmetries of the boundary theory in flat space. 
On the subsector, one can write the supergravity potential describing
RG flows through steepest descent in the
canonical form.
From the effective non-trivial scalar potential (\ref{V}) which consists of
two parts, one expects 
that the superpotential we are considering maybe encoded in either 
$A_1^{IJ}$ tensor  or $A_{2, L}^{\;\;\;\;\;IJK}$ tensor. It turns out that 
one of the eigenvalues of $A_1^{\;\;IJ}$ tensor (\ref{a1}), $z_3$,
provides a ``superpotential'' $W$ related to
scalar potential $V$ by
\bea 
V(\la, \la',\al, \phi)= g^2 \left[ \frac{16}{3} \left|\frac{\partial
z_3}{\partial \la} \right|^2 + 4 \left|\frac{\partial z_3}{\partial
\la'} \right|^2 - 6  \left|z_3\right|^2 \right], 
\label{pot} 
\eea
where $z_3$ is a function of $\la, \la', \al$ and $\phi$:
\bea 
z_3(\la, \la',\al, \phi) & = & 6e^{i(\alpha + 2\phi)}p^2qr^2t^2
 + 6e^{2i(\alpha + \phi)}pq^2r^2t^2  + p^3(r^4 + e^{4i\phi}t^4)  \nonu \\
& & +
e^{3i\alpha}q^3(r^4 + e^{4i\phi}t^4), 
\label{superpotential}
\eea
with (\ref{pqrt}).
At first sight, there is 
no dependence on the derivatives of $z_3$ with respect to
the fields $\al$ and $\phi$ in the (\ref{pot}).
We have found that the complex-valued superpotential $z_3$ satisfies the 
following algebraic relations\footnote{This observation was motivated by
the result of \cite{ar2} where only a single relation holds because $G_2$
invariant sector satisfies $\la=\la'$ and $\al=\phi$. Similar aspect 
happens for $SO(3)$-invariant 
$AdS_5$ gauged supergravity in \cite{pilchwarner}.}:
\bea
\partial_{\al} \log \left|z_3\right| 
& = &
2\sqrt{2} \;p \;q \;\partial_{\la} \mbox{Arg} z_3^{\ast}, 
\nonu \\ 
\partial_{\phi} \log \left|z_3\right| & = &
2\sqrt{2} \;r \;t \;\partial_{\la'} \mbox{Arg} z_3^{\ast},
\label{identity}
\eea
which relate the derivative of magnitude of $z_3$ with respect to 
$\al(\phi)$ to the one of angle of $z_3^{\ast}$ with respect to $\la(\la')$.
Then it is elementary to show that  
one can express the scalar potential by exploiting the above 
relations as following form indicating the magnitude of
$z_3$ serves as the true superpotential: 
\bea
W(\la, \la', \al, \phi)  & = &  |z_3|, \nonu \\
V(\la, \la',\al, \phi) & = & g^2 \left[  \frac{16}{3} \left(\partial_{\la}
W \right)^2  + \frac{2}{3p^2 q^2}
\left(\partial_{\al}
W \right)^2 + 4 
\left(\partial_{\la'} 
W  \right)^2+ \frac{1}{2r^2 t^2} \left(\partial_{\phi} 
W \right)^2 - 6  W^2 \right].
\label{potandsuper}
\eea
Let us note that by differentiating this $V$ with respect to
one of fields among $\la, \la',\al$ and $ \phi$, the scalar potential
$V$ has critical points at 1) critical 
points of $W$ and at 2) 
points for which $W$ satisfies some differential equation.
In this sense, the role of superpotential $W$ is important because 
the property of critical points of scalar potential is encoded in those
of superpotential.
At the three 
supersymmetric critical points(Table 1) in the supergravity context 
for which supersymmetric flows
are generically simpler, more controllable and represent very stable fixed 
points than nonsupersymmetric ones, the absolute values of
gradients of $z_3$ with respect to
$\la, \la'$ vanish. That is $\left| \frac{\pa z_3}{\pa \la}\right| =
\left| \frac{\pa z_3}{\pa \la'}\right|=0$. 
 In other words, in terms of $W$, they are equivalent to  
$\partial_{\la} W = \partial_{\la'} W =
\partial_{\al} W  = \partial_{\phi} W= 0$. 
This implies that supersymmetry preserving vacua have negative cosmological
constant:the scalar potential $V$  at the three critical points becomes
$V=-6 g^2 |z_3|^2$ or $ W =\sqrt{-V/6g^2}$.
The critical points of $W$ 
yield supersymmetric 
stable $AdS_4$ vacua in supergravity which will 
 imply non-trivial conformal fixed points in the dual field theory under 
appropriate conditions. 
Supersymmetry ensures that there are no unstable modes
in a supersymmetry preserving solution to the supergravity
equations. 
The other critical points of $V$ yield 
nonsupersymmetric(but usually $AdS_4$) vacua that may or may not
be stable(in order to be stable, the small oscillations must satisfy the
Breitenlohner-Freedman condition \cite{bf}).
The superpotential $W$ has the 
following values at the various supersymmetric or non-supersymmetric
critical points. There is well-known trivial critical point, 
corresponding to the $\S^7$ compactification of the 11-dimensional
supergravity, at which
all the supergravity 
scalar and pseuo-scalar fields vanish and whose cosmological
constant is $\La = - 6 g^2$ and which preserves ${\cal N}=8$ supersymmetry.

\bea
\begin{array}{|c|c|c|c|}
\hline $\mbox{Gauge symmetry}$ & s, s', \al, \phi   & W & V \nonu \\
\hline
   SO(8) 
& s = 0 = s'
  & 1
 & -6 g^2 \nonu \\
\hline
   SO(7)^{-}  
& s=\pm \frac{1}{2}, s'=\frac{1}{2}, \al=\frac{\pi}{2} =\phi
  & \frac{3 \times 5^{3/4}}{8}  &- \frac{25\sqrt{5}}{8}g^2  \nonu \\
\hline
   SO(7)^{+}    & 
 s=\sqrt{\frac{1}{2}(\frac{3}{\sqrt{5}}-1)} =s', 
 \al=0=\phi
 & \frac{3}{2} \times 5^{-1/8} & - 2\times5^{3/4}g^2 \nonu \\
\hline
   G_{2}   
&  s =\pm \sqrt{\frac{2}{5}(\sqrt{3}-1)}, s'=\sqrt{\frac{2}{5}(\sqrt{3}-1)},
&  & \nonu \\
& \al = \cos^{-1} \frac{\sqrt{3-\sqrt{3}}}{2} =\phi 
& \sqrt{ \frac{36\sqrt{2} \times 3^{1/4}}{25\sqrt{5}}} 
& - \frac{216\sqrt{2} \times 3^{1/4}}{25\sqrt{5}}g^2
\nonu \\ \hline
   SU(4)^{-}    
&  s=0, s'=1, \phi=\frac{\pi}{2}   & \frac{3}{2}  & 
- 8g^2 \nonu \\ \hline
 SU(3) \times U(1)    
&  s=\frac{1}{\sqrt{3}}, s'=\frac{1}{\sqrt{2}}, \al=0,
\phi=\frac{\pi}{2}  & \frac{3^{3/4}}{2} & - \frac{9 \sqrt{3}}{2} g^2 \nonu
\\ \hline
\end{array}
\nonu
\eea
Table 1. \sl Summary of various critical points in the context of
superpotential : symmetry group,
vacuum expectation values of fields, superpotential and 
cosmological constants. We have taken the first, second and fourth columns 
from \cite{warner}. \rm

$\bullet$ $SO(8)$ case: ${\cal N} =8$

At this point, complex self-dual tensor 
$\phi_{ijkl}$ vanishes from (\ref{phiijkl}) because $\la$ and $\la'$ vanish. 
In the dual field theory 70 scalars
are mapped into relevant chiral primary operators. The ${\bf 35_{v}}$ scalars
correspond to conformal dimension of 1, $\mbox{Tr} X^i X^j -\frac{1}{8}
\de^{ij} \mbox{Tr} X^2$ where $X^i$ is an eight scalars ${\bf 8_v}$ 
of $SO(8)$ of 3-dimensional ${\cal N}=8$ $SU(N_c)$ gauge theory, 
after dualizing
the gauge field,
 while   the ${\bf 35_{c}}$ pseudo-scalars
correspond to conformal dimension of 2, 
 $\mbox{Tr} \la^i \la^j -\frac{1}{8}
\de^{ij} \mbox{Tr} \la^2$ where $\la^i$ is an eight fermion fields ${\bf 8_c}$. 
The RG trajectories of the relevant operators will interpolate 
the ${\cal N}=8$ $SO(8)$ fixed point to other {\it new} fixed points if
the supergravity potential allows additional stable critical points besides
the $SO(8)$ invariant point at  $\phi_{ijkl}=0$.

$\bullet$ $SO(7)^{-}$ case: ${\cal N} =0$

In this case, all the eigenvalues of  $A_1^{\;\;IJ}$ tensor (\ref{a1}), 
$z_1, z_2$ and $z_3$ are complex and equal and their magnitude is
given in Table 1. 
It is known that the number of 
supersymmetries is equal to the number of eigenvalues of 
 $A_1^{\;\;IJ}$ tensor whose absolute value are the same as 
$\sqrt{-\Lambda/6 g^2}$. It is easy to check that there is no supersymmetry and
the lack of supersymmetry makes it hard to verify the results
in the dual field theory.  
This critical point was found in \cite{englert} and is unstable and this fact
suggests that the IR field theory limit maybe non-unitary.
Since $\al=\pi/2=\phi$, this corresponds to giving only the pseudo-scalars
 expectation values corresponding to non-zero internal  magnetic 
{\it four-form} field strength \cite{englert} in $d=11$.

$\bullet$ $SO(7)^{+}$ case: ${\cal N} =0$

In this case, also all the eigenvalues of  $A_1^{\;\;IJ}$ 
tensor are real and equal and given in Table 1.  
Since $\al=0=\phi$, this corresponds to giving only the scalars
expectation values corresponding to some perturbation of the {\it metric
} tensor in a dimensional reduction by some twisted $\S^7$.  
This critical point was found in \cite{dn} and is unstable and there is no
corresponding $SO(7)^{+}$ invariant $d=3$ CFT.

$\bullet$ $G_2$ case: ${\cal N} =1$

The eigenvalue $z_2$ is equal to $z_1$ which is different
from $z_3$.  
So there exist two eigenvalues with degeneracies 7, 1 \cite{ar2}. 
Since the absolute value of $z_3$(nothing but the superpotential) 
is the same as 
$\sqrt{-\Lambda/6 g^2}$(see Table 1), 
this gives rise to ${\cal N} =1$ supersymmetry
that is a degeneracy of $|z_3| =W$.
Simultaneously turning on both scalars and pseudo-scalars, one gets
this vacuum. Group theoretically it is impossible to break $SO(8)$ 
into $G_2$ by giving expectation values to fields in a single ${\bf 35_v}$
or ${\bf 35_c}$
of $SO(8)$. The analysis of superpotential and scalar potential in these 
three cases was already
given in \cite{ar2} and one can see, by putting $\la=\la'$ and $\al=\phi$
in the (\ref{superpotential}), that it will lead to the one  given in 
\cite{ar2}. Our $z_3$ corresponds to their $z_2$. The scalar potential
reduces to 
\bea
V(\al, \la)_{SO(7)^{\pm}, G_2}=
2 g^2 \left( (7 v^4-7 v^2 +3) c^3 s^4 +(4v^2-7) v^5 s^7 + c^5 s^2 +
7 v^3 c^2 s^5 -3 c^3 \right),
\nonu
\eea
together with (\ref{cs}).
Although the complete spectrum at the IR fixed point is not known,
the chiral operators may be followed since their dimensions are protected from
quantum corrections. 

$\bullet$ $SU(4)^{-}$ case: ${\cal N} =0$

All the eigenvalues of $A_1^{\;\;IJ}$ tensor (\ref{a1}) are equal and given in
Table 1. 
Since $\la=0$ and $\phi=\pi/2$, this invariant critical point occurs
at  purely pseudo-scalar expectation values and was found in \cite{pw}. 
This can be seen by breaking
$SO(7)^{-}$ of the first vacuum into $SO(6)^{-}=SU(4)^{-}$ which is contained in
$SO(7)^{-}$. In this case,
28-beins $u^{IJ}_{\;\;\;\;KL}$ and $v^{IJKL}$ are expressed in compact form
as 
\bea
u^{IJ}_{\;\;\;\;KL}  = \frac{1}{4} \left(
( c'+ 1) \de^{IJ}_{\;\;\;\;KL} + (c'-1) 
F_{\;\left[K \right.}^{-\left[I \right.} 
F_{\;\left. L \right]}^{-\left. J \right]} \right), \qquad 
v^{IJKL} & = & -\frac{1}{2} s' \; Y^{2-}_{IJKL},
\nonu
\eea
where $
F_{\;I}^{-J} =  \mbox{diag} (if,if,if,-if)$ and $ f =
\left(\begin{array}{cc} 
0 & 1 \nonu \\ 
-1 &  0
\end{array} \right) $
with (\ref{Y}) and (\ref{cs}). Moreover, the scalar potential can be
written as 
$
V(\la')_{SU(4)^{-}}= 2 g^2 \left( s'^4-2s'^2-3 \right).
\nonu
$
The stability of critical point is not known.
In 11-dimensional supergravity theory \cite{pw}, the metric on the $\S^7$
is distorted by stretching the $U(1)$ fibers and four-form field strength
is nonzero in the $\S^7$ direction.

$\bullet$ $SU(3) \times U(1)$ case: ${\cal N} =2$

The eigenvalue $z_2$ is equal to $z_3$ and given in Table 1
which is different
from $z_1(=2/3^{1/4})$. So there are 
two eigenvalues with degeneracies 6, 2 \cite{ap}.
By putting $\al=0$ and $\phi=\pi/2$
in  (\ref{superpotential}), it will lead to the one  given in 
\cite{ap}. Our $z_3$ corresponds to their $z_2$.
In this case, the scalar potential can be expressed as
$
V(\la, \la')_{SU(3) \times U(1)}=
 2 g^2 c'^2 \left( \left(s^3+c^3 \right) s'^2 -3c \right)
\nonu
$.
The critical point may be thought of as IR fixed point of the dual field
theories on the branes. Since all the cosmological constants are negative
and admit $AdS_4$ metrics, the corresponding gauge theories are conformal.
Note that superpotential $W$ becomes real and this fact made it easier to find a BPS 
domain-wall solutions. Existence of an algebraic 
identity (\ref{identity}) may reflect that the supersymmetry restricts the 
structure on the scalar sectors but this is too sufficient condition since
$SU(3) \times U(1)$ critical point 
does not possess that kind of identity. Maybe 
the group theoretical structure of $G_2$ symmetry rather than supersymmetry 
alone restricts the behavior of superpotential.

Let us begin with the resulting Lagrangian of the 
scalar-gravity sector by explicitly finding out the scalar kinetic terms
appearing in the action (\ref{action}) in terms of $\la, \la', \al$ and 
$\phi$. 
By taking the product of $  A^{\;\;IJKL}_{\mu}$ appearing (\ref{kinsu3}) and
its complex conjugation and taking into account the 
multiplicity four(for given index pairs, there are four possible choices),  
we arrive at the following expression
\bea 
\left| A^{\;\;IJKL}_{\mu} \right|^2 
 =  36 \left( \left(
\partial_{\mu} \lambda \right)^2 + 2 s^2 \left(
\partial_{\mu} \alpha \right)^2 \right) + 
48 \left( \left( \partial_{\mu} \lambda' \right)^2 +
2 s'^2 \left( \partial_{\mu} \phi \right)^2 \right).
\nonu
\eea
In old days, the significance of construction of kinetic terms was not 
emphasized because, at that time, they concerned about only the structure
of extrema of scalar potential. As we mentioned ealier, our approach 
to get kinetic terms directly through 28-beins is more appropriate. 
Therefore the resulting Lagrangian of scalar-gravity sector takes the form:
\bea
\int d^4 x \sqrt{-g} \left( \frac{1}{2} R  
- \frac{3}{8}\left(
\partial_{\mu} \lambda \right)^2 -\frac{3}{4} s^2 \left(
\partial_{\mu} \alpha \right)^2  -
\frac{1}{2}  \left( \partial_{\mu} \lambda' \right)^2 -
 s'^2 \left( \partial_{\mu} \phi \right)^2  - 
V(\la, \la', \al, \phi) \right),
\label{action1}
\eea
together with (\ref{cs}), (\ref{pqrt}) and (\ref{potential}).

Having established the holographic duals of both supergravity critical points,
and examined small perturbations around the corresponding fixed point
field theories, one can proceed the supergravity description of the
 RG flow between the two fixed points.
The supergravity scalars whose vacuum expectation values lead to the new 
critical point tell us what relevant operators in the dual field theory 
would drive a flow to the fixed point in the IR. 
To construct the superkink(providing for a geometric description of 
RG flows) corresponding  to 
the supergravity description of the nonconformal
RG flow from one scale to other two 
 connecting critical points in $d=3$ conformal field theories,
the form of a 3d Poincare invariant metric but breaking 
the full conformal group $SO(3,2)$ invariance takes the form:
\bea
ds^2= e^{2A(r)} \eta_{\mu \nu} dx^{\mu} dx^{\nu} + dr^2, \;\;\;
\eta_{\mu \nu}=(-,+,+),
\label{ansatz}
\eea
characteristic of space-time with a domain wall where $r$ is the
coordinate transverse to the wall(can be interpreted
 as an energy scale) and $A(r)$ is the scale factor
in the four-dimensional metric.

By change of variable $U(r)=e^{A(r)}$ at the critical points,
the geometry becomes $AdS_4$ space with a cosmological constant
$\La$ equal to the value of $V$ at the critical points: 
$\La= -3 (\pa_r A)^2$. In the dual theory, this corresponds to  
a superconformal fixed point of the RG flow(from one scale to another).
Our interest in domain wall space-times comes from
their connection to the RG flow of the dual field theories.
The variable $U$, the distance from the horizon, 
can be identified with RG scale and linearly proportional to
the energy scale of the boundary theory which is an important 
aspect of the AdS/CFT correspondence. $U=\infty$ corresponds
to long distance in the bulk(UV in the dual field theory) and $U=0$(near
$AdS_4$ horizon
corresponds to short distances in the bulk(IR in the dual field theory).
This implies that the RG flow of the coupling constants  of the field theory
is encoded in the $U$ dependence of the supergravity scalar fields.
At a fixed point the scalar field is constant and therefore corresponding 
$\be$-function vanishes. 
We are looking for ``interpolating'' solutions that are asymptotic
to $AdS_4$ space both for $\la \rightarrow \la_{UV}, \la' \rightarrow
\la'_{UV}, \al \rightarrow \al_{UV}, \phi \rightarrow \phi_{UV}$ 
for $r \rightarrow 
\infty $ so that the background is asymptotic to the   
supersymmetric $AdS_4$ background at infinity 
while $\la \rightarrow \la_{IR}, 
\la' \rightarrow \la'_{IR}, \al \rightarrow \al_{IR}, \phi \rightarrow 
\phi_{IR}$ for
$r \rightarrow -\infty$ and so we approach a {\it new} conformal fixed point.
The $AdS_4$ geometries at the endpoints imply conformal symmetry in the
UV and IR limits of the field theory and there exists $OSp(8|4)$ symmetry
at the UV fixed point while $OSp({\cal N}|4)$ symmetry at the IR end.
We will show how supergravity can provide a description of the entire
RG flow from the maximal supersymmetric UV theory to the lower IR fixed point.
With the above  ansatz (\ref{ansatz})
the equations of motion for the scalars and the metric from (\ref{action1}) 
read
\bea 
& &
4 \pa_r^2 A + 6 (\pa_r A)^2 +
\frac{3}{4} (\pa_r \la)^2   +
\frac{3}{2} s^2 (\pa_r \alpha)^2  + 
(\partial_r \la')^2 + 2 s'^2 ( \pa_r \phi )^2
  + 2V = 0, \nonu \\ 
& & \pa_r^2 \la +
3 \pa_r A \pa_r \la  - 
 \sqrt{2} s c ( \partial_r \alpha )^2 - \frac{4}{3} \pa_{\la} V   
=0, \nonu \\ 
& &
\pa_r^2 \la' +
3 \pa_r A \pa_r \la'  - 
\sqrt{2} s' c' ( \partial_r \phi )^2 -  \pa_{\la'} V = 0, 
\nonu \\ 
& &
s^2 \pa_r^2 \al + 3s^2 \pa_r A \pa_r \al + \sqrt{2} s c \pa_r \al
\pa_r \la -\frac{2}{3} \pa_{\al} V = 0,
\nonu \\ 
&&
s'^2 \pa_r^2 \phi + 3s'^2 \pa_r A \pa_r \phi + \sqrt{2} s' c' \pa_r \phi
\pa_r \la' -\frac{1}{2} \pa_{\phi} V = 0.
\label{eom}
\eea 

By substituting the domain-wall ansatz (\ref{ansatz})
into the Lagrangian (\ref{action1}), 
the Euler-Lagrangian equations are the second, third, fourth and fifth 
equations of
(\ref{eom}) for the functional $E[A, \la, \la', \al, \phi]$ 
\cite{st} with the integration by parts
on the term of $\pa_r^2 A $.
The energy-density per unit area transverse to $r$-direction is given by
\bea
 E[A,\la,\la',\alpha,\phi]&  = & \int_{-\infty}^{\infty} dr e^{3A}
\left[- 3 \left( 2 (\pa_r A )^2 + \pa_r^2 A \right) -3 \left(
 \frac{1}{8}
\left( \partial_r \la \right)^2 +
p^{2}q^{2}\left( \partial_r \alpha \right)^2
 \right) \right.   \nonu \\ 
& & \left. -\frac{1}{2}\left(
 \left( \partial_r \la' \right)^2 +
8 r^{2} t^{2}\left( \partial_r \phi \right)^2
 \right) -V(\la, \la', \al, \phi)\right].
\nonu
\eea
We are looking for a nontrivial configuration along $r$-direction and in order
to find out the first-order differential equations the domain-wall satisfy,
let us rewrite and reorganize
 the energy-density by sum of complete squares plus others due to
usual squaring-procedure as follows:
\bea
&& E[A,\la,\la',\alpha,\phi] =  \nonu \\ 
& &    -\frac{1}{2} \int_{-\infty}^{\infty} dr e^{3A}
\left[ - 6 \left( \pa_r A + \sqrt{2} g \left|z_3\right|
\right)^2 + \frac{3}{4} \left|  \pa_r \la  - 
i 2 \sqrt{2} p q \partial_r \alpha - 
\frac{8\sqrt{2}}{3} g   e^{2i\beta} \partial_{\la} z_3
  \right|^2  \right. \nonu
\\ &  &  
+ \left|  \pa_r \la'  - 
i 2 \sqrt{2} r t \partial_r \phi - 
2 \sqrt{2} g   e^{2i\beta} \partial_{\la'} z_3
  \right|^2
+ 12\sqrt{2} g \pa_r A \left| z_3 \right| \nonu \\ 
& &   +
2\sqrt{2}\left( \partial_r \la  - i
2\sqrt{2} p q \partial_r \alpha \right) 
g e^{-2i\beta} \partial_{\la} z_3^{\ast} 
  + 
2\sqrt{2}\left( \partial_r \la  + i
2\sqrt{2} p q \partial_r \alpha \right) 
g e^{2i\beta} \partial_{\la} z_3 
\nonu \\ & & \left. +
2\sqrt{2}\left( \partial_r \la'  - i
2\sqrt{2} r t \partial_r \phi \right) 
g e^{-2i\beta} \partial_{\la'} z_3^{\ast} 
  + 
2\sqrt{2}\left( \partial_r \la  + i
2\sqrt{2} r t \partial_r \phi \right) 
g e^{2i\beta} \partial_{\la'} z_3 
\right],  
\nonu
\eea
where $z_3^{\ast} =|z_3| e^{2i \be}$.
Then one can easily check that
the last eight cross-terms in the above can be 
expressed as
$4 \sqrt{2} g e^{3A} \pa_r |z_3| $ by using previous remarkable 
identities (\ref{identity}). Therefore one arrives at  
\bea
& &  -\frac{1}{2}
\int_{-\infty}^{\infty} dr e^{3A} \left[ 
 - 6 \left( \pa_r A + \sqrt{2} g \left|z_3\right|
\right)^2 + \frac{3}{4} \left|  \pa_r \la  - 
i 2 \sqrt{2} p q \partial_r \alpha - 
\frac{8\sqrt{2}}{3} g   e^{2i\beta} \partial_{\la} z_3
  \right|^2  \right. \nonu
\\ &  & \left. 
+ \left|  \pa_r \la'  - 
i 2 \sqrt{2} r t \partial_r \phi - 
2 \sqrt{2} g   e^{2i\beta} \partial_{\la'} z_3
  \right|^2 \right]  - 2\sqrt{2}g \left( 
e^{3A}\left|z_3\right|\right) |_{- \infty}^
{\infty}. 
\nonu
\eea
Finally, we find BPS bound, inequality of the energy-density
\bea
E[A,\la,\la',\alpha,\phi]
 \geq  - 2\sqrt{2}g \left( e^{3A(\infty)} W(\infty) -
e^{3A(-\infty)}W(-\infty)\right). 
\label{Ebound}
\eea 

Then $E[A, \la, \la', \al, \phi]$ is extremized by the following
so-called BPS domain-wall solutions. The first order differential
equations for the scalar fields are the gradient flow equations
of a superpotential defined on a restricted four-dimensional slice of
the scalar manifold and simply related to the potential of gauged supergravity
on this slice via (\ref{potandsuper}).
\bea  
\partial_{r}\la & = & \pm
\frac{4\sqrt{2}}{3}g\left(e^{-2i\beta}\frac{\partial z_3^{\ast}}{\partial
\la} + e^{2i\beta}\frac{\partial z_3}{\partial \la}\right)=
\pm \frac{8\sqrt{2}}{3}g \partial_{\la} W ,\nonu
\\
\partial_{r}\la' & = & \pm
\sqrt{2}g\left(e^{-2i\beta}\frac{\partial z_3^{\ast}}{\partial \la' } +
e^{2i\beta}\frac{\partial z_3}{\partial \la'}\right)=
\pm 2 \sqrt{2} g \pa_{\la'} W, \nonu \\
\partial_{r}\alpha & = & \mp \frac{2}{3pq} i g \left(e^{-2i\beta}\frac{\partial
z_3^{\ast}}{\partial\la} - e^{2i\beta}\frac{\partial
z_3}{\partial\la}\right) = \pm
\frac{\sqrt{2}}{3p^2 q^2} g \pa_{\al} W, \nonu \\ 
\partial_{r}\phi & = & \mp
\frac{1}{2rt} i g \left(e^{-2i\beta}\frac{\partial
z_3^{\ast}}{\partial\la'} - e^{2i\beta}\frac{\partial z_3}{\partial\la'}
\right) = \pm \frac{\sqrt{2}}{3r^2 t^2} g \pa_{\phi} W , \nonu \\ 
\partial_{r}A & = & \mp \sqrt{2} g W.
\label{first}
\eea
It is evident that the left hand sides of the first four relations 
vanish as one approaches the supersymmetric extrema, i.e. 
$\partial_{\la} W = \partial_{\la'} W =
\partial_{\al} W = \partial_{\phi} W= 0$
thus indicating a domain-wall configuration that is a topological soliton
with a nontrivial kink number along $r$-direction.
The asymptotic behaviors of $A(r)$ are $A(r) \rightarrow r/r_{UV} + const$
for $r \rightarrow \infty$ and  
$A(r) \rightarrow r/r_{IR} + const$
for $r \rightarrow -\infty$. Then by differentiating $A(r)$ wih respect to
$r$, those of $\pa_r A$ become 
$\pa_r A \rightarrow 1/r_{UV}$
for $r \rightarrow \infty$ and  
$\pa_r A \rightarrow 1/r_{IR}$ for $r \rightarrow -\infty $.
At the two critical points, since $V =-6 g^2 W^2$,  
one can write 
the inverse radii of $AdS_4$ as cosmological constant or superpotential $W$.
Therefore we conclude that $1/r$ is equal to $\pm \sqrt{2} g 
W$. This fact
is encoded in the last equation of (\ref{first}).
It is straightforward to verify that any solutions 
$\{\la(r), \la'(r), \al(r),\phi(r), A(r)\}$ of (\ref{first}) 
satisfy the gravitational and
scalar equations of motion given by the
second order differential equations (\ref{eom}).
Embedding or consistent truncation means that
the flow is entirely determined by the
equations of motion of supergravity in four-dimensions and any solution
of the truncated theory can be lifted to a solution of untruncated 
theory \cite{dn86}.
Using (\ref{first}), the monotonicity \cite{domainwall} of 
$\pa_r A $  which is related to the local potential energy
of the superkink leads to
\bea
\pa_r^2 A  = -2 g^2 \left( \frac{8}{3} \left( \partial_{\la} W
 \right)^2 +2 \left( \partial_{\la'}  W
\right)^2  +\frac{1}{3 p^2 q^2} \left( \partial_{\al}  W
\right)^2  + \frac{1}{3 r^2 t^2} \left( \partial_{\phi}  W
\right)^2 \right)  \leq 0.
\nonu
\eea
Note that the value of superpotential at either end of a kink may be thought
of as determining the topological sector. 

One can understand the above bound (\ref{Ebound}) 
as a conseqence of supersymmetry preserving
bosonic background. 
In order to find supersymmetric bosonic backgrounds, the variations
of spin-$1/2$ and spin-$3/2$- fields should vanish. 
From \cite{dn2}, the gravitational and scalar parts of these variations are:
\bea
\de\psi_{\mu}^i & = & 
 2 D_{\mu} \epsilon^i - \sqrt{2} g A_1^{\;ij} \ga_{\mu} \ep_j, 
\nonu \\
\de \chi^{ijk}  & = & - \ga^{\mu} A_{\mu}^{\;\;ijkl} \ep_l- 2 g 
A_{2l}^{\;\;\;ijk} 
\ep^l,
\label{susy}
\eea
where the covariant derivative acting on supersymmetry parameter is  
\bea
 D_{\mu} \epsilon^i = \partial_{\mu} \ep^i - \frac{1}{2} 
\omega_{\mu a b} \sigma^{ab} \ep^i
+ \frac{1}{2} 
{\cal B}_{\mu\;\;\;j}^{\;\;i} \ep^j, \;\;\; 
{\cal B}_{\mu\;\;\;j}^{\;\;i} = \frac{2}{3} \left( u^{ik}_{\;\;\;IJ}
\partial_{\mu} u_{jk}^{\;\;\;IJ} - v^{ikIJ} \partial_{\mu} v_{jkIJ} \right).
\label{cov}
\eea
Here $\ep_i$ and $\ep^j$ are complex conjugates each other under the chiral 
basis\footnote{
In this basis, the $\gamma$ matrices satisfy $\{\gamma^{\mu}, 
\gamma^{\nu} \} = 2 g^{\mu \nu}$ and  $\gamma^i =
\left(
\begin{array}{cc}
0 & -\sigma^i \nonu \\
\sigma^i & 0 \\
\end{array}
\right)$ where $\sigma^i$ are Pauli matrices and
$\gamma^5 = i \gamma^0 \gamma^1 \gamma^2 \gamma^3$. }. 
The field ${\cal B}_{\mu\;\;\;j}^{\;\;i}$ is a $SU(8)$ gauge field for a local
$SU(8)$ invariance, $\omega$ a spin connection, $\sigma$ a commutator
of two $\gamma$ matrices. 
Under the projection operators $(1\pm \gamma_5)/2$ where 
$
\gamma_5=\left(
\begin{array}{cc}
1 & 0 \nonu \\
0 & -1 \\
\end{array}
\right),
$
the supersymmetry parameter 
$\ep_i$ has the four column components as $(\eta_1, \eta_2,
0, 0)$ where $\eta_1, \eta_2$ are complex spinor fields.
Moreover, complex conjugate 
$\ep^i$ is the charge conjugate spinor of $\ep_i$ and satisfies
$\ep^i = C {\gamma^0}^{T} {\ep_i}^{\ast}$ where $C=
\left(
\begin{array}{cc}
\si^2 & 0 \nonu \\
0 & -\si^2 \\
\end{array}
\right)
$ 
we introduce has the 4 column components as $(0, 0, \eta_3=i {\eta_2}^{\ast}, 
\eta_4=-i{\eta_1}^{\ast})$.

The variation of 56 Majorana spinors $\chi^{ijk}$ gives rise to
the first order differential equation of $\la, \la', \al$ and 
$\phi$ by exploiting
the explicit forms of $ A_{\mu}^{\;\;ijkl}$ (\ref{kinsu3}) and 
$A_{2l}^{\;\;\;ijk}$ (\ref{su3y}) in the
appendix.
Although there is a summation over the last index $l$ appearing in
$ A_{\mu}^{\;\;ijkl}$ and $A_{2l}^{\;\;\;ijk}$ in the right hand side of
(\ref{susy}), the structure of them implies
that summation runs over only one index. For example, when $i=1, j=7, k=2$ 
and $l=8$, the vanishing of variation of  $\chi^{ijk}$ leads to 
\bea
\left(  \partial_r \la +i 2 \sqrt{2} p q \partial_r \al \right) \gamma^2
\ep_8=
 \frac{8\sqrt{2} g}{3} \frac{\partial z_3^{\ast} }{\partial \la } \ep^8, 
\nonu
\eea  
and its complex conjugation. We used the fact that
$y_3$ is proportional to $e^{i \al} \frac{\pa z_3^{\ast}}{\pa \la}$ according
to (\ref{su3y}):
this functional relation implies that the scalar potential can be expressed
in terms of $z_3$ and is expressed in terms of $z_3$ and $ \pa_{\la} z_3$.  
Recognizing that $ \gamma^2 \ep_8 ={\ep^8}^{\ast}$,
we arrive at
\bea
\left(  \partial_r \la +i 2 \sqrt{2} p q \partial_r \al \right) =
 \frac{8 \sqrt{2} g}{3} \frac{\partial z_3^{\ast}}{\partial \la } \left(
\frac{\ep^8}{|\ep^8|} \right)^2.
\label{diff}
\eea
Therefore one obtains two relations for $\la$ and $\al$ fields from this and
its complex conjugation:
\bea
\partial_r \la = \frac{4\sqrt{2}}{3} g \left(e^{-2 i \be}
\frac{\partial z_3^{\ast}}{\partial \la } + e^{2 i \be} \frac{\partial 
z_3}
{\partial \la} \right), \qquad
\partial_r \al = -\frac{2}{3 p q}i g \left(e^{-2 i \be}
\frac{\partial z_3^{\ast}}{\partial \la } - e^{2 i \be} 
\frac{\partial z_3}
{\partial \la} \right),
\nonu
\eea
where complex spinor field has a phase $\be(r)$ and
$\eta_3 =|\eta_3(r)| e^{i \beta(r)}$. This is nothing but
the first and third equations of (\ref{first}).
It is straightforward to re-express them in terms of a derivative
of $W$ with respect to $\la$ field by writing $z_3^{\ast} 
= W e^{2i \be}$.
On the other hand, when $i=6, j=2, k=4$ 
and $l=8$, the vanishing of variation of  $\chi^{ijk}$ leads to 
\bea
\left(  \partial_r \la' +i 2 \sqrt{2} r t\partial_r \phi \right) \gamma^2
\ep_8=
 \frac{8\sqrt{2} g}{3} \frac{\partial z_3^{\ast} }{\partial \la' } \ep^8, 
\nonu
\eea  
and its complex conjugation. Again, we used 
the fact that $y_7$ is proportional to $e^{i \phi} \frac{\pa z_3^{\ast}}{\pa \la'}$
from (\ref{su3y}).
It also implies that the scalar potential can be expressed
in terms of $z_3$ and is expressed in terms of $z_3 $ and $ \pa_{\la'} z_3$. 
Therefore one obtains the following relations for $\la'$ and $\phi$
fields
\bea
\partial_r \la' = \sqrt{2} g \left(e^{-2 i \be}
\frac{\partial z_3^{\ast}}{\partial \la' } + e^{2 i 
\be} \frac{\partial z_3}
{\partial \la'} \right), \qquad
\partial_r \phi = -\frac{1}{2 r t}i g \left(e^{-2 i \be}
\frac{\partial z_3}{\partial \la' } - e^{2 i \be} \frac{\partial z_3^{\ast}}
{\partial \la'} \right).
\nonu
\eea
which are exactly the same as the second and fourth equations of (\ref{first})
leading to a derivative of superpotential with respect to $\la'$ field as 
before.


Putting it in another way, the cross terms for second equation of (\ref{susy}) 
can be simplified by using 
the identity of \cite{dn} 
\bea
D_{\mu}  A_1^{\;ij} = \frac{\sqrt{2}}{24} \left( A_{2\;klm}^{\;\;i} 
A_{\mu}^{\;\;jklm} + A_{2\;klm}^{\;\;j} 
A_{\mu}^{\;\;iklm} \right), 
\nonu
\eea
and by realizing the following identity 
\bea
 A_1^{\;ik}  A_{1\;kj} -\frac{1}{18} A_{2\;klm}^{\;\;i}
A_{2j}^{\;\;\;klm} = -\frac{1}{6 g^2} V \de^i_j,  
\nonu
\eea
the spin-$1/2$ variations vanish if and only if 
the steepest descent equations given by first four
equations of (\ref{first}) are satisfied. 

Moreover, the variation of gravitinos 
$\psi_{\mu=1}^i$ 
will leads to
$
i \eta_3 \left( \partial_r A + \sqrt{2} g z_3^{\ast} e^{-2 i \beta} \right) =0.
$ 
Similar relation for spinor field $\eta_4$ holds. 
By realizing $\partial_r A$ is real, one can conclude that
$e^{-2 i \beta}=-|z_3|/z_3^{\ast}$.
Finally, we obtains
\bea
 \partial_r A = - \sqrt{2} g W,
\nonu
\eea
which is the same as the last equation of (\ref{first}).
Similar equation appears in the $\eta_4$ 
spinor component. There are no 
other additional equations for $\mu=0, 3$ indices.

According to (\ref{cov}), $\omega_{\mu=2, a, b}$ term has nonvanishing
$\omega_{\mu=2, 1, 1}$ but these are summed over 
$\si^{1,1}=[\ga^1, \ga^1]$ which is identically zero. 
Therefore there is no contribution on this
part:
\bea
2 \partial_r   \ep^8+ \frac{i}{2} \left[ 3(-1+c) \pa_{r} \alpha +4
\left(-1+c' \right) \pa_{r} \phi \right] \ep^8 - \sqrt{2} g z_3
\left(
\begin{array}{cccc}
0 &0 & 0 &i \nonu \\
0 & 0 &-i &0 \nonu \\
0 &-i &0 &0 \nonu \\
i &0 &0 &0 \nonu
\end{array}
\right)
\left(
\begin{array}c
\eta_1 \nonu \\
\eta_2 \nonu \\
0 \nonu \\
0 \nonu
\end{array}
\right) = 0,
\nonu
\eea   
where we used the fact that  
the $SU(8)$ connection ${\cal B}^{\;\;I}_{\mu \;\;J}$ 
defined by (\ref{cov}) obtained 
by plugging (\ref{uvsu3}) has the following diagonal
form:
\bea
{\cal B}^{\;\;I}_{\mu \;\;J} & = & \mbox{diag} \left(-i q^2 
\partial_{\mu} \alpha, -i q^2 
\partial_{\mu} \alpha, -i q^2 
\partial_{\mu} \alpha, -i q^2 
\partial_{\mu} \alpha, -i q^2 
\partial_{\mu} \alpha, -i q^2 
\partial_{\mu} \alpha, \right. \nonu \\
& & \left. \frac{i}{2} \left[ 3(-1+c) \pa_{\mu} \alpha -4
\left(-1+c' \right) \pa_{\mu} \phi \right],
\frac{i}{2} \left[ 3(-1+c) \pa_{\mu} \alpha +4
\left(-1+c' \right) \pa_{\mu} \phi \right] \right),  
\nonu
\eea
together with (\ref{pqrt}) and (\ref{cs}).
Finally, one of the variation of gravitinos $\psi_{\mu=2}^i$ gives rise to
\bea
2 \partial_r \eta_3 + \frac{i}{2} \left[ 3(-1+c) \pa_r \alpha +4
\left(-1+c' \right) \pa_r \phi \right] \eta_3 - \sqrt{2} g z_3^{\ast} 
{\eta_3}^{\ast} =0. 
\nonu
\eea
From this,
we get two relations for spinor field $\eta_3$, and
using $\eta_3 =|\eta_3(r)| e^{i \beta(r)}$ and plugging 
back,
we get
\bea
\partial_r \beta = \frac{1}{4} \left[ 3(-1+c) \pa_r \alpha +4
\left(-1+c' \right) \pa_r \phi \right], \qquad
\partial_r |\eta_3| = \frac{\sqrt{2}}{2} g W |\eta_3| e^{-2 i \beta}.
\nonu
\eea

One can show that there exists a supersymmetric flow if and only if 
the equations (\ref{first}) are satisfied, that is, the flow is determined
by the steepest descent of the superpotential and the cosmology $A(r)$ is
determined directly from this steepest descent.

Let us consider mass, $\widetilde{M^2}$ for the $\overline{\la},
\overline{\la'}, \overline{\al}$ and $\overline{\phi} $ at the
critical points of superpotential $W$ where 
$\overline{\la}=\sqrt{\frac{3}{4}}
\la,
\overline{\la'}=\la', \overline{\al}=\sqrt{\frac{3}{2}} \al$ and 
$\overline{\phi}=\sqrt{2} \phi $. 
By differentiating (\ref{pot}) and putting
$\partial_{\la} W = \partial_{\la'} W =
\partial_{\al} W  = \partial_{\phi} W= 0$, we get
\bea
\widetilde{M}^2_{ij}   = \pa_{\phi_i} \pa_{\phi_j} V=  2 g^2 W^2
 {\cal U}_{ik}  \left( {\cal U}_{kj} -3 \de_{kj} \right), \qquad
\phi_i = (\overline{\la}, \overline{\la'}, \overline{\al}, 
\overline{\phi} ),   \nonu
\eea
where $\cal U$ is related to the second derivatives of $W$ with respect to
various fields. 
The mass scale is set by the inverse radius, $1/r$, of the $AdS_4$ space
and this can be written as $1/r=\ell_p \sqrt{-V/3}=\sqrt{2} g 
 W  $ where we used $V=-6g^2  W^2$. Via AdS/CFT
correspondence, ${\cal U}$ is related to the conformal dimension $\De$ of
the field theory operator dual to the fluctuation of the fields 
$\overline{\la},
\overline{\la'}, \overline{\al}$ and $\overline{\phi} $. 
Since the matrix ${\cal U}$ is real
and symmetric, it has real eigenvalues $\de_k$ and the eigenvalues of
$\widetilde{M}^2 r^2 $ are given by $\de_k \left( \de_k -3 \right)$. 
Since a new radial coordinate $U(r)=e^{A(r)}$ is 
the renormalization group scale on the flow, we should
find the leading
contributions to the $\be$ functions of the couplings $\overline{\la},
\overline{\la'}, \overline{\al}$ and $\overline{\phi} $
in the neighborhood of the end points of the flow. 
At a fixed point, the fields are constant and corresponding $\beta$ function
vanishes.
Since $\frac{d}{d r}=
\frac{d A}{d r} U \frac{d}{d U}=-\sqrt{2} g W U \frac{d}{d U}$, (\ref{first})
becomes 
$
U \frac{d}{d U} \phi_i  =  - \frac{2}{W}  \frac{\partial W}{  
\partial \phi_i}  \approx - 
 {\cal U}_{i j}|_{\mbox{crit.pt.}}
\de \phi_j, 
$
where we expanded to the first order in the neighborhood of a critical point.
Thus ${\cal U}$ determines the behavior of the $\overline{\la},
\overline{\la'}, \overline{\al}$ and $\overline{\phi} $
near the critical points. 
The RG flow of the coupling constants of the field theory is encoded 
in the $U$ dependence of the fields. 
To depart the UV fixed point($U=+ \infty$), the 
flow must take place in directions in which the operators must be relevant,
and to approach the IR fixed point($U \rightarrow 0$), the corresponding 
operators must be irrelevant.  

\subsection{$SO(5)$ Sector }

$\bullet$ $SO(5)^{+}$ embedding

The superpotential $W$ is generically extracted as an eigenvalue
of the $A_1^{\;\;IJ}$ tensor from (\ref{z1so5+}) and it is related to the 
scalar potential as follows:
\bea 
V(\la, \mu,\rho) & = & g^2 \left[\frac{32}{5} \left( \partial_{\la}
W \right)^2 + \frac{32}{5}\left( \partial_{\mu} W \right)^2 + 
\frac{32}{5}\left( \partial_{\rho} W \right)^2 \right. \nonu \\ 
& & \left. -\frac{16}{5}
\partial_{\la} W \partial_{\mu} W  -\frac{16}{5}   
\partial_{\la} W \partial_{\rho} W
  -\frac{16}{5}   \partial_{\mu} W  
\partial_{\rho} W  - 6  W^2 \right], 
\nonu
\eea
where the superpotential is a real function of $\la, \mu$ and $\rho$ 
\bea 
W(\la, \mu, \rho)  = -\frac{1}{8\sqrt{uvw}} \left(5 + u^2v^2 + 
\mbox{two cyclic permutations}
 \right).
\nonu
\eea
There is a trivial critical point at which
all the fields vanish and whose cosmological
constant is $\La = - 6 g^2$ preserving ${\cal N}=8$ supersymmetry.

\bea
\begin{array}{|c|c|c|c|}
\hline $\mbox{Gauge symmetry}$ & \la, \mu, \rho   & W & V \nonu \\
\hline
   SO(8), {\cal N}=8 & \la=\mu=\rho=0 &  -1  & -6 g^2 \nonu \\
\hline
   SO(7)^{+}, {\cal N}=0 & \la=\mu=-\rho= \frac{\sqrt{2}}{4} \log 5 &  
-\frac{3}{2 \times 5^{1/8}} & - 2\times 5^{3/4} g^2 \nonu \\
\hline
\end{array}
\nonu
\eea
Table 2. \sl Summary of one critical point in the context of
superpotential : symmetry group,
vacuum expectation values of fields, superpotential and 
cosmological constants. \rm

In this case, there exists an unstable  nonsupersymmetric critical point with
$SO(7)^{+}$ gauge symmetry.  
By taking the product of $  A^{\;\;IJKL}_{\mu}$ (\ref{kineticso5}) and
its complex conjugation, $ A_{\mu \;IJKL}$, 
and summing over all the indices with appropriate multiplicities, 
we arrive at the following expression
\bea 
\left| A^{\;\;IJKL}_{\mu} \right|^2 
 =  
144 \left( \left(
\partial_{\mu} \lambda \right)^2 + \left(\pa_{\mu} \mu \right)^2
+ \left( \pa_{\mu} \rho \right)^2 \right) + 
96 \left( \partial_{\mu} \lambda \pa_{\mu} \mu  +
\pa_{\mu} \la \pa_{\mu } \rho + \pa_{\mu} \mu \pa_{\mu} \rho \right).
\nonu
\eea
By substituting the domain-wall ansatz (\ref{ansatz}) as before into 
the resulting Lagrangian of the scalar and gravity
part we have not presented here explicitly and 
by plugging the above kinetic terms, one gets the Euler-Lagrangian 
equations. Along the flow between $SO(8)$ fixed point and $SO(7)^{+}$
fixed point we are considering, the relations 
$\mu=-\rho =\la$ hold. Therefore, after repeating the analysis of the 
energy-density,
the first order differential equations
for the $\la$ field are the gradient flow equations of corresponding
superpotential defined on a single dimensional slice of the scalar manifold:   
\bea
\partial_{r} \la  =  \pm
\frac{8\sqrt{2}}{7}g \partial_{\la} W,\qquad
\partial_{r}A  =  \mp \sqrt{2} g W,
\nonu
\eea
which is also obtained by putting $\al=0=\phi$ and $\la'=\la$
in the previous subsection for $SU(3)$ invariant sectors. 
Unlike as supersymmetric flow we have studied in $SU(3)$ invariant sectors,
at both ends, the derivative of $W$ with respect to $\la$ does not vanish:
at supersymmetric $SO(8)$ fixed point, it vanishes, while, at 
nonsupersymmetric $SO(7)^{+}$ fixed point, it does not.

$\bullet$ $SO(5)^{-}$ embedding

The scalar potential can be written in terms of superpotential as follows:
\bea 
V(\la, \mu,\rho) & = & g^2 \left[\frac{32}{5} \left| \partial_{\la}
W \right|^2 + \frac{32}{5}\left| \partial_{\mu} W \right|^2 + 
\frac{32}{5}\left| \partial_{\rho} W \right|^2 \right. \nonu \\ 
& &  -\frac{8}{5}
\partial_{\la} W \partial_{\mu} W^{\ast}
-\frac{8}{5}
\partial_{\mu} W \partial_{\la} W^{\ast}
  -\frac{8}{5}   
\partial_{\la} W \partial_{\rho} W^{\ast}
 -\frac{8}{5}   
\partial_{\rho} W \partial_{\la} W^{\ast} \nonu \\
& & \left.
  -\frac{8}{5}   \partial_{\mu} W  
\partial_{\rho} W^{\ast} -\frac{8}{5}   \partial_{\rho} W  
\partial_{\mu} W^{\ast}  - 6  \left|W \right|^2 \right], 
\nonu
\eea
where the complex-superpotential from (\ref{z1so5-}) takes the form: 
\bea
W(\la,\mu,\rho) = \frac{(1+i)}{16(uvw)^{3/2}} \left(
 -i u^2 + u^3 v^3 w +\mbox{two cyclic permutations}+ 5 uvw-5 i u^2v^2w^2  
\right). 
\nonu
\eea
The superpotential $W$ has the following values at two nonsupersymmetric 
critical points besides the supersymmetric one.
Like as nonsupersymmetric flow for $SO(5)^{+}$ embedding case,
at supersymmetric fixed point, derivatives of superpotential 
with respect to fields vanish while, at 
nonsupersymmetric fixed points, they do not vanish.

\bea
\begin{array}{|c|c|c|c|}
\hline $\mbox{Gauge symmetry}$ & \la, \mu, \rho   & W & V \nonu \\
\hline
   SO(8), {\cal N}=8  & \la=\mu=\rho=0 &  1  & -6 g^2 \nonu \\
\hline
   SO(7)^{-}, {\cal N}= 0
 & \la=\mu=-\rho= \sqrt{2} \log \frac{1+\sqrt{5}}{2} &  
\frac{3}{8} \sqrt{11+2i} &  - \frac{25\sqrt{5}}{8}g^2 \nonu \\
\hline
   SO(6)^{-}, {\cal N}=0  & \la=\sqrt{2} \log (\sqrt{2}+1), \mu=0,
\rho= \sqrt{2} \log (\sqrt{2}-1) &  
\frac{3}{2} & - 8 g^2 \nonu \\
\hline
\end{array}
\nonu
\eea
Table 3. \sl Summary of various critical points in the context of
superpotential : symmetry group,
vacuum expectation values of fields, superpotential and 
cosmological constants. \rm

Contrary to the previous $SO(5)^{+}$ embedding case, 
there are no such first order differential equations for either a flow between
$SO(8)$ fixed point and $SO(7)^{-}$ fixed point or a flow between 
SO(8) fixed point and $SO(6)^{-}$ fixed point. Furthermore,
in \cite{romans} there were no extrema with gauge symmetry of 
$SO(5), SO(5) \times U(1)$ or $SO(5) \times SU(2)$ which leads to 
the fact that the effective four-dimensional theory of Awada et al 
\cite{awada} is {\it not} a sector of de Wit-Nicolai theory. 
It can be a sector of the full $d=4$  theory resulting from compactification
on $\S^7$ in which some of the massive scalars are given expectation values.

\subsection{$SO(3) \times SO(3)$ Sectors }

During the flow connecting $SO(8)$ fixed point to 
 $SO(3) \times SO(3)$ fixed point, the six fields $\la^{\al}$ vanish 
for $\al=2, \cdots, 7$. We are considering domain-walls in supergravity with
a nontrivial superpotential defined on a restricted one-dimensional
slice of the scalar manifold.
One of the eigenvalues of $A_1^{\;\;IJ}$ tensor, $z_7$ (\ref{z1so3so3})
restricted on  $\la^{\al}, \al=2, \cdots, 7$
provides a ``superpotential'' $W$ related to
scalar potential $V$ by
\bea
V(\la^1)|_ {\la^2= \cdots = \la^7=0} &  = & 
\frac{g^2}{16} \left(-61 -36 \cosh \la^1 + \cosh 2\la^1 \right)
 \nonu \\
& = & 
 g^2 \left[ 8 
\left( \pa_{\la^1} W \right)^2 - 6 W^2 \right],
\label{potentialso3so3}
\eea
where the superpotential is 
$
W(\la^1) = \frac{1}{4} \left( 3 + \cosh \la^1 \right).
$
The plots of $V$ and $W$ on the $\la^1$ parameter space are shown 
in Figure 1. 

\begin{figure}[htb]
\label{fig1}
\vspace{0.5cm}
\epsfysize=7cm
\epsfxsize=7cm
\centerline{
\epsffile{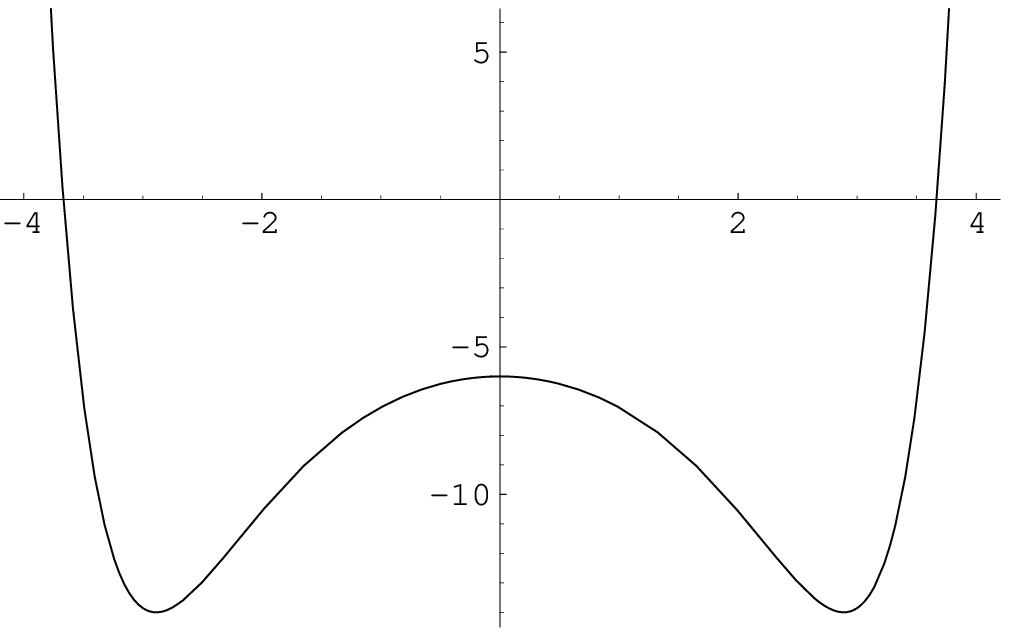}  
\epsfysize=7cm
\epsfxsize=7cm
\epsffile{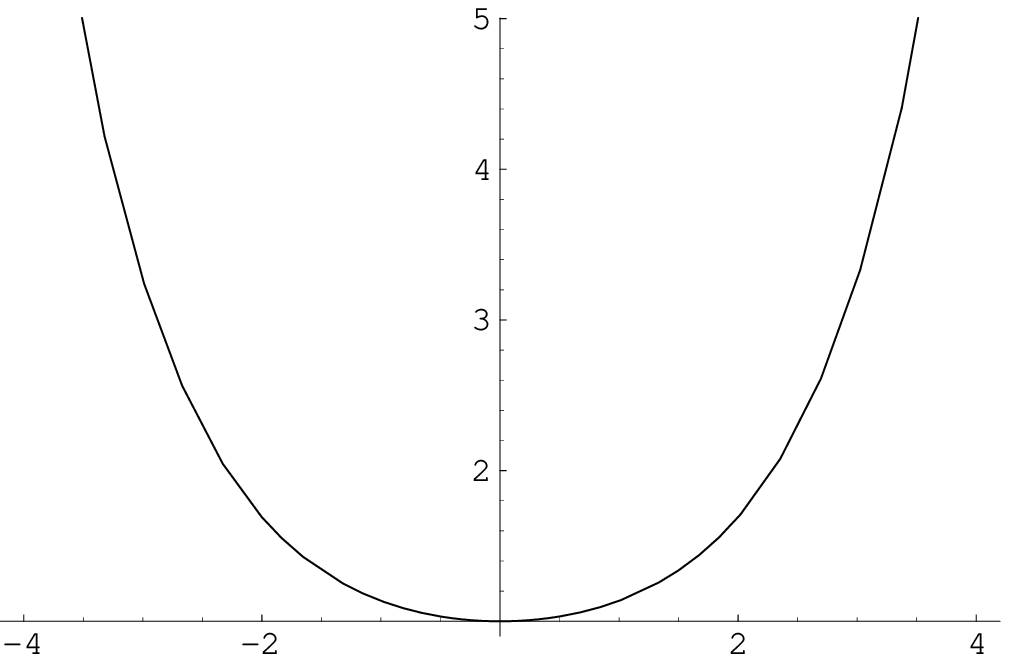}  }
\vspace{0.5cm}
\caption{\sl 
The plots of $V$ (on the left) and $W$ (on the right), with
$\la^1$ on the horizontal axis.  Scalar potential $V$, 
at $\la^1=0$, is the maximally supersymmetric  and locally maximum 
while superpotential $W$ at that point is locally minimum. The cosmological 
constant is $-6$. We took $g^2$ as 1 for simplicity. 
At $\la^1=2 \sinh^{-1} 2=2.89$, $V$ has locally minimum and 
is nonsupersymmetric
and the cosmological 
constant is $-14$. See also \cite{distler}.}
\end{figure}
\vspace{0.5cm}

At the supersymmetric $SO(8)$ fixed point, 
the critical point of scalar potential
$V$ is nothing but the one of superpotential $W$ while, at the 
nonsupersymmetric $SO(3) \times SO(3)$, the critical point of $V$ 
is not a critical point of $W$(that is, $\pa_{\la^1} W$ does not vanish 
at a point) but at point for which $W$ satisfies
$4 \pa^2_{\la^1} W -3 W =0$(note that, at a fixed point, $\cosh \la^1 =9$) 
if we differentiate $V$ with respect to
$\la^1$.  

\bea
\begin{array}{|c|c|c|c|}
\hline $\mbox{Gauge symmetry}$ & \la^{\al}   & W & V \nonu \\
\hline
   SO(8), {\cal N}=8 
& \la^{\al} = 0,  1 \leq \al \leq 7 &  1  & -6 g^2 \nonu \\
\hline
   SO(3) \times SO(3), {\cal N}=0  & \sinh \left(\frac{\la^1}{2} \right) = 2,
\la^{\al} = 0,  2 \leq \al \leq 7 &  3  & - 14 g^2 \nonu \\
\hline
\end{array}
\nonu
\eea
Table 4. \sl Summary of one critical point in the context of
superpotential : symmetry group,
vacuum expectation values of fields, superpotential and 
cosmological constants. \rm

The resulting Lagrangian of the scalar-gravity sector after
finding out the kinetic terms  according to (\ref{so3so3kinetic}) 
and by realizing from correct counting of multiplicities that
$
|A^{\;\;IJKL}_{\mu}|^2 = 6 (1+\cosh\la^1)^2 (\pa_{\mu} \la^1)^2
$
in terms of $\la_1$ takes the form with (\ref{potentialso3so3}):
\bea
\int d^4 x \sqrt{-g} \left( \frac{1}{2} R  
- \frac{1}{16} \left(1+\cosh \la^1 \right)^2 \left(
\partial_{\mu} \lambda^1 \right)^2   - 
V(\la^1) \right).
\nonu
\eea
By substituting the domain-wall ansatz (\ref{ansatz})
into this Lagrangian, one obtains
domain-wall solutions by direct minimization of energy-functional 
when we assume quadratic order in the fluctuation of
$\la_1$
\bea
\partial_{r} \la^1  =  \pm
4\sqrt{2} g \partial_{\la^1} W,\qquad
\partial_{r}A  =  \mp \sqrt{2} g W.
\nonu
\eea
Although the right hand side of the first relation vanishes at the supersymmetric
$SO(8)$ fixed point, for the nonsupersymmetric $SO(3) \times SO(3)$ fixed 
point, the ``velocity'' of $\la^1$ does not vanish because the right hand side
$\pa_{\la^1} W $ at that point has nonzero value.
The analytic solutions for these become
\bea
\la^1(r) & = & 
\log \frac{1+e^{\sqrt{2} g ( c-r)}}{1-e^{\sqrt{2} g ( c-r)}}, \qquad
A(r)  =  \frac{1}{4} \left( 3 \sqrt{2} g r + \log [2 
\sinh{\sqrt{2} g (c-r)}]\right). \nonu
\eea
Although it is not known whether the Breitlohner-Freedman condition 
is satisfied by the solution in the ${\cal N}=8$ theory, the solution 
is {\it stable} in the context of positive energy theorem 
without supersymmetry 
\cite{boucher}. So, in the terminology of \cite{st}, this solution is 
``non-BPS'' domain-wall solution interpolating between supersymmetric 
$SO(8)$ vacuum and nonsupersymmetric $SO(3) \times SO(3)$ one.
 
\newpage

\section{Appendix A: $SU(3)$ Invariant Sectors}

The 28-beins $u^{IJ}_{\;\;\;KL}$ and $v^{IJKL}$ 
fields which are elements of $56 \times 56$ ${\cal V}(x)$ 
(\ref{56bein}) of
the fundamental 56-dimensional representation of $E_7$
 can be obtained 
by exponentiating the vacuum expectation values $\phi_{ijkl}$ 
(\ref{phiijkl}) of $SU(3)$ singlet space via (\ref{calV}). After tedious 
calculation,  the nonzero components of the 28-beins have the
following seven $4 \times 4$ block diagonal matrices respectively:
\bea 
u^{IJ}_{\;\;\;KL} & = & \mbox{diag}
(u_1,u_2,u_3,u_4,u_5,u_6,u_7), \nonu \\
v^{IJKL} & = &
\mbox{diag} (v_1,v_2,v_3,v_4,v_5,v_6,v_7). \nonu 
\eea 
Each hermitian(for example, 
$(u_1)^{78}_{\;\;\;12}=((u_1)^{12}_{\;\;\;78})^{\ast}=BK^2$) 
submatrix is $4 \times 4$ matrix and we denote
antisymmetric index pairs $[IJ]$ and $[KL]$ explicitly for convenience.
For simplicity, we make an empty space corresponding to lower triangle 
elements:
\bea 
&& u_1
= \left(
\begin{array}{ccccc}
 & \left[12 \right] & \left[34 \right] & \left[56 \right] &
\left[78 \right] \nonu \\
 \left[12 \right] &  A & B & B &  \frac{B}{K^2} \nonu \\
\left[34 \right]  &  & A & B & \frac{B}{K^2} \nonu \\
 \left[ 56\right] &  &  & A &
\frac{B}{K^2} \nonu \\
 \left[ 78\right] &  &   &  & A \\
\end{array}
\right), 
u_2 = \left(
\begin{array}{ccccc}
& \left[13 \right] & \left[24 \right] & \left[57 \right] &
\left[68 \right] \nonu \\
 \left[13 \right] & C & -D & -\frac{L E}{K} &  \frac{E}{K L} \nonu \\
\left[24 \right]  &  & C &\frac{L E}{K}  & -
\frac{E}{K L} \nonu \\
 \left[ 57\right] &  &  & C & - 
\frac{D}{L^2} \nonu \\
 \left[ 68\right] &  &   &  & C \\
\end{array}
\right), \nonu \\ && u_3 = \left(
\begin{array}{ccccc}
& \left[14 \right] & \left[23 \right] & \left[58 \right] &
\left[67 \right] \nonu \\
 \left[14 \right] & C & D & \frac{E}{K L} &  \frac{L E}{K} \nonu \\
\left[23 \right]  &  & C & \frac{E}{K L} &
\frac{L E}{K} \nonu \\
 \left[ 58 \right] &  &  
& C & L^2 D \nonu \\
 \left[ 67 \right] &  &
&   & C \\
\end{array}
\right), u_4 = \left(
\begin{array}{ccccc}
& \left[15 \right] & \left[26 \right] & \left[37 \right] &
\left[48 \right] \nonu \\
 \left[15 \right] & C & -D & \frac{L E}{K} & - 
\frac{E}{K L} \nonu \\
\left[26 \right]  &  & C & -\frac{L E}{K} &
\frac{E}{K L} \nonu \\
 \left[37\right] & & & C & - \frac{D}{L^2} 
\nonu \\
 \left[48 \right] & &  &  & C \\
\end{array}
\right), \nonu \\ && u_5 = \left(
\begin{array}{ccccc}
& \left[16 \right] & \left[25 \right] & \left[38 \right] &
\left[47 \right] \nonu \\
 \left[16 \right] & C & D & - \frac{E}{K L} &  - 
\frac{L E}{K} \nonu \\
\left[25 \right]  &  & C & - \frac{E}{K L} &
-\frac{L E}{K} \nonu \\
 \left[38 \right] &  &  & C & L^2 D \nonu \\
 \left[47 \right] &  &  
&   & C \\
\end{array}
\right), u_6 = \left(
\begin{array}{ccccc}
& \left[17 \right] & \left[28 \right] & \left[35 \right] &
\left[46 \right] \nonu \\
 \left[17 \right] & C & -\frac{D}{L^2} & -\frac{K E}{L} & 
\frac{K E}{L} \nonu \\
\left[28 \right]  &  & C & K L E & -
K L E \nonu \\
 \left[35 \right] &  &  & C & -D \nonu \\
 \left[46 \right] &  &  &  & C \\
\end{array}
\right), \nonu \\ && u_7 = \left(
\begin{array}{ccccc}
& \left[18 \right] & \left[27 \right] & \left[36 \right] &
\left[45 \right] \nonu \\
 \left[18 \right] & C & L^2 D & K L E & 
K L E \nonu \\
\left[27 \right]  &  & C & \frac{K E}{L} &
\frac{K E}{L} \nonu \\
 \left[36 \right] &  &  & C & D \nonu \\
 \left[45 \right] &  &  &  & C
\end{array} \right), 
 v_1 = \left(
\begin{array}{ccccc}
& \left[12 \right] & \left[34 \right] & \left[56 \right] &
\left[78 \right] \nonu \\
 \left[12 \right] & \frac{F}{K} & \frac{G}{K} & \frac{G}{K} 
& K G \nonu \\
\left[34 \right]  & & \frac{F}{K} & \frac{G}{K}
& K G \nonu \\
 \left[ 56\right] &  &  & \frac{F}{K} &
K G \nonu \\
 \left[ 78\right] &  &  &  & 
K^3 F \\
\end{array}
\right), \nonu \\ & & v_2 = \left(
\begin{array}{ccccc}
& \left[13 \right] & \left[24 \right] & \left[57 \right] &
\left[68 \right] \nonu \\
 \left[13 \right] & \frac{H}{K} & -\frac{I}{K} & -\frac{J}{L} 
&  L J \nonu \\
\left[24 \right]  &  & \frac{H}{K} & \frac{J}{L} &
-L J \nonu
\\
 \left[ 57\right] &  &  & 
\frac{K H}{L^2} 
& -K I \nonu \\
 \left[ 68\right] &  &  &  & K L^2 H \\
\end{array}
\right), v_3 = \left(
\begin{array}{ccccc}
& \left[14 \right] & \left[23 \right] & \left[58 \right] &
\left[67 \right] \nonu \\
 \left[14 \right] & \frac{H}{K} &  \frac{I}{K} & L J 
&  \frac{J}{L} \nonu \\
\left[23 \right]  &   &  \frac{H}{K} & L J &
\frac{J}{L} \nonu \\
 \left[ 58 \right] &  &  & K L^2 H &  K I \nonu \\
 \left[ 67 \right] &  &  &   & 
\frac{K H}{L^2} \\
\end{array}
\right), \nonu \\ & & v_4 = \left(
\begin{array}{ccccc}
& \left[15 \right] & \left[26 \right] & \left[37 \right] &
\left[48 \right] \nonu \\
 \left[15 \right] & \frac{H}{K} & -\frac{I}{K} & 
\frac{J}{L} &  -L J \nonu \\
\left[26 \right]  &  & \frac{H}{K} & -\frac{J}{L}
& L J \nonu
\\
 \left[37\right] &  &  & \frac{K H}{L^2} & -K I \nonu \\
 \left[48 \right] &  &   &  & K L^2 H \\
\end{array}
\right), v_5 = \left(
\begin{array}{ccccc}
& \left[16 \right] & \left[25 \right] & \left[38 \right] &
\left[47 \right] \nonu \\
 \left[16 \right] & \frac{H}{K} & \frac{I}{K} & -L J 
&  -\frac{J}{L} \nonu \\
\left[25 \right]  &  & \frac{H}{K} & -L J &
-\frac{J}{L} \nonu
\\
 \left[38 \right] & &  & K L^2 H & K I \nonu \\
 \left[47 \right] &  &  &   & 
\frac{K H}{L^2} \\
\end{array}
\right), \nonu \\ & & v_6 = \left(
\begin{array}{ccccc}
& \left[17 \right] & \left[28 \right] & \left[35 \right] &
\left[46 \right] \nonu \\
 \left[17 \right] & \frac{K H}{L^2} & -K I & 
-\frac{J}{L} & \frac{J}{L} \nonu \\
\left[28 \right]  &  & K L^2 H &
L J & -L J \nonu \\
 \left[35 \right] &  & & \frac{H}{K} & 
-\frac{I}{K} \nonu \\
 \left[46 \right] &  &  &  & 
\frac{H}{K} \\
\end{array}
\right), v_7 = \left(
\begin{array}{ccccc}
& \left[18 \right] & \left[27 \right] & \left[36 \right] &
\left[45 \right] \nonu \\
 \left[18 \right] & K L^2 H & K I & 
L J & L J \nonu \\
\left[27 \right]  &  & \frac{K H}{L^2} &
\frac{J}{L} & \frac{J}{L} \nonu \\
 \left[36 \right] &  & & \frac{H}{K} & 
\frac{I}{K} \nonu \\
 \left[45 \right] &  &   & 
& \frac{H}{K} \\
\end{array} \right), \\
\label{uvsu3} 
\eea 
where simplified quantities are functions of $\la, \la', \al$
and $\phi$  
\bea 
& & A \equiv p^3, \;\;\; B \equiv pq^2,
\;\;\; C \equiv pr^2, \;\;\; D \equiv pt^2, \;\;\;E \equiv qrt
\;\;\; F \equiv q^3,\nonu \\ && G \equiv p^2q, \;\;\;H \equiv
qt^2, \;\;\; I \; \equiv qr^2, \;\;\;J \equiv prt 
\;\;\; K \equiv e^{i \al}, \;\;\; L \equiv e^{i \phi} 
\nonu
\eea 
and $p,q, r$ and $t$ are some functions of $\la$ and $\la'$ in (\ref{pqrt}).
The lower triangle part can be read off from the upper triangle part by
hermitian property.
Also, 28-beins $u_{IJ}^{\;\;\;KL}$ and $v_{IJKL}$ are obtained by taking a
complex conjugation of (\ref{uvsu3}).
The nonzero components of $A_2$ tensor, $A_{2, L}^{\;\;\;\;\;IJK}$, 
are obtained from
(\ref{ttensor}) and (\ref{a1a2}) by simply plugging (\ref{uvsu3}) into there 
and 
they are classified by eight different fields with degeneracies 12, 3, 3,
12, 12, 4, 4, 6 respectively  and  
given by:
\bea
& & A_{2, 1}^{\;\;\;\;\;\;256}=A_{2, 1}^{\;\;\;\;\;234}= A_{2,
2}^{\;\;\;\;\;165}= A_{2, 2}^{\;\;\;\;\;\;143}=A_{2,
3}^{\;\;\;\;\;456}= A_{2, 3}^{\;\;\;\;\;412}= A_{2,
4}^{\;\;\;\;\;\;365}=A_{2, 4}^{\;\;\;\;\;321}= \nonu \\ && A_{2,
5}^{\;\;\;\;\;\;634}=A_{2, 5}^{\;\;\;\;\;612}= A_{2,
6}^{\;\;\;\;\;543}= A_{2, 6}^{\;\;\;\;\;\;521}\equiv y_1,\nonu \\
& & A_{2, 7}^{\;\;\;\;\;128}= A_{2, 7}^{\;\;\;\;\;348}= A_{2,
7}^{\;\;\;\;\;\;568} \equiv y_2,  \nonu \\
&&A_{2,
8}^{\;\;\;\;\;172}=A_{2, 8}^{\;\;\;\;\;\;437}=A_{2,
8}^{\;\;\;\;\;657}\equiv y_3, \nonu \\ && A_{2,
1}^{\;\;\;\;\;375}= A_{2, 1}^{\;\;\;\;\;\;674}=A_{2,
2}^{\;\;\;\;\;574}= A_{2, 2}^{\;\;\;\;\;673}= A_{2,
3}^{\;\;\;\;\;\;571}=A_{2, 3}^{\;\;\;\;\;276}=A_{2,
4}^{\;\;\;\;\;\;275}=A_{2, 4}^{\;\;\;\;\;176}= \nonu \\ &&  A_{2,
5}^{\;\;\;\;\;247}= A_{2, 5}^{\;\;\;\;\;\;173}=A_{2,
6}^{\;\;\;\;\;237}= A_{2, 6}^{\;\;\;\;\;147}\equiv y_4,\nonu \\ &&
A_{2, 1}^{\;\;\;\;\;\;368}=A_{2, 1}^{\;\;\;\;\;458}=A_{2,
2}^{\;\;\;\;\;\;358}=A_{2, 2}^{\;\;\;\;\;648}= A_{2,
3}^{\;\;\;\;\;618}= A_{2, 3}^{\;\;\;\;\;\;528}=A_{2,
4}^{\;\;\;\;\;\;826}=A_{2, 4}^{\;\;\;\;\;518}= \nonu \\ & & A_{2,
5}^{\;\;\;\;\;814}= A_{2, 5}^{\;\;\;\;\;\;823}=A_{2,
6}^{\;\;\;\;\;813}= A_{2, 6}^{\;\;\;\;\;428}\equiv y_5,\nonu \\ &&
A_{2, 7}^{\;\;\;\;\;\;513}=A_{2, 7}^{\;\;\;\;\;326}=A_{2,
7}^{\;\;\;\;\;\;416}=A_{2, 7}^{\;\;\;\;\;425}\equiv y_6, \nonu \\
&& A_{2, 8}^{\;\;\;\;\;624}= A_{2, 8}^{\;\;\;\;\;\;415}=A_{2,
8}^{\;\;\;\;\;316}= A_{2, 8}^{\;\;\;\;\;325}\equiv y_7, \nonu \\
&& A_{2, 1}^{\;\;\;\;\;\;278}=A_{2, 2}^{\;\;\;\;\;718}=A_{2,
3}^{\;\;\;\;\;\;478}=A_{2, 4}^{\;\;\;\;\;738}=A_{2,
5}^{\;\;\;\;\;\;578}=A_{2, 6}^{\;\;\;\;\;758}\equiv y_8, \nonu
\eea 
where their explicit forms are   
\begin{eqnarray}
 y_1 & = &- e^{-2i(\alpha +
\phi)} \left(e^{3i\alpha}p^2qr^2t^2 + e^{i(3\alpha + 4\phi)}p^2qr^2t^2 +
pq^2r^2t^2 + e^{4i\alpha}pq^2r^2t^2 \right. \nonu \\ 
& & \left. + e^{i(\alpha +
2\phi)}q(2q^2r^2t^2 + p^2(r^2 + t^2)^2) + e^{2i(\alpha +
\phi)}p(2p^2r^2t^2 + q^2(r^2 + t^2)^2) \right), \nonu \\ 
y_2 & = & -
e^{-2i\alpha} \left(e^{3i\alpha}p^2qr^4 + 2e^{i(\alpha + 2\phi)}q(2p^2 +
q^2)r^2t^2 + 2e^{2i(\alpha + \phi)}p(p^2 + 2q^2)r^2t^2 \right.
\nonu \\ 
& & \left. + pq^2r^4 + e^{i(3\alpha + 4\phi)}p^2qt^4 + 
e^{4i\phi}pq^2t^4 \right)
\nonu \\
& = &  {\it \bf
- \frac{2\sqrt{2}}{3} e^{i \al} \frac{\pa z_2^{\ast}}{\pa \la}}, \nonu \\
y_3 & = & - e^{-2i(\alpha + 2\phi)} \left(e^{i(3\alpha +
4\phi)}p^2qr^4 + e^{4i\phi}pq^2r^4 + 2e^{i(\alpha + 2\phi)}q(2p^2
+ q^2)r^2t^2 \right. \nonu \\
 & & \left.+ 2e^{2i(\alpha + \phi)}p(p^2 + 2q^2)r^2t^2 + e^{3i\alpha}p^2qt^4 +
 pq^2t^4 \right)  \nonu \\
&=& {\it \bf - \frac{2\sqrt{2}}{3} e^{i\al} 
\frac{\pa z_3^{\ast}}{\pa \la}}, \nonu \\ 
y_4 & = & - e^{-i(\alpha + 3\phi)}rt \left(e^{4i\phi}p^2qr^2
 + e^{i(3\alpha + 4\phi)}pq^2r^2 + p^2qt^2 + e^{3i\alpha}pq^2t^2 \right.
\nonu \\
 & &  \left. + e^{2i(\alpha + \phi)}q(2p^2 + q^2)(r^2 + t^2) + e^{i(\alpha +
 2\phi)}p(p^2 + 2q^2)(r^2 + t^2) \right), \nonu \\ 
y_5 & = & - e^{-i(\alpha + \phi)}rt \left(p^2qr^2
 + e^{3i\alpha}pq^2r^2 + e^{4i\phi}p^2qt^2 + e^{i(3\alpha + 
4\phi)}pq^2t^2 \right. \nonu \\
 & &  \left.+ e^{2i(\alpha + \phi)}q(2p^2 + q^2)(r^2 + t^2) + e^{i(\alpha +
 2\phi)}p(p^2 + 2q^2)(r^2 + t^2) \right), \nonu \\ 
y_6 & = & - e^{-i(3\alpha + 
\phi)} \left(e^{i\alpha}p
 + q \right) rt \left(
e^{2i\alpha}p^2r^2 - e^{i\alpha}pqr^2 + q^2r^2 + e^{2i(\alpha + 
2\phi)}p^2t^2 \right. \nonu \\
 & & \left. - e^{i(\alpha + 4\phi)}pqt^2 + e^{4i\phi}q^2t^2 + 3e^{i(\alpha +
 2\phi)}pq(r^2 + t^2) \right)  \nonu \\
& = & {\it \bf - \frac{\sqrt{2}}{2} e^{-i \phi} 
\frac{\pa z_2^{\ast}}{\pa \la'}},  \nonu \\
y_7 & = & - e^{-3i(\alpha + \phi)} \left(
e^{i\alpha}p
 + q \right) rt 
\left(e^{2i(\alpha + 2\phi)}p^2r^2 - e^{i(\alpha + 4\phi)}pqr^2 + 
e^{4i\alpha}q^2r^2 \right. \nonu \\
 & &  \left.+ e^{2i\alpha}p^2t^2 - e^{i\alpha}pqt^2 + q^2t^2 + 3e^{i(\alpha +
 2\phi)}pq(r^2 + t^2) \right) \nonu \\
&=& {\it \bf - \frac{\sqrt{2}}{2} e^{i \phi} 
\frac{\pa z_3^{\ast}}{\pa \la'}}, \nonu \\  
y_8 & = & - e^{-2i\phi} \left(p +
 e^{i\alpha}q \right) \left(p^2r^2t^2
 + e^{4i\phi}p^2q^2t^2 - e^{i\alpha}pqr^2t^2 - e^{i(\alpha + 
4\phi)}pqr^2t^2 \right. \nonu \\
 & &  \left. + e^{2i\alpha}q^2r^2t^2 + 
e^{2i(\alpha + 2\phi)}q^2r^2t^2 + e^{i(\alpha +
 2\phi)}pq(r^4 + 4r^2t^2 + t^4) \right).
\label{su3y}
\end{eqnarray}
Notice that we did not
write down the components of $A_2$ tensor which are interchanged
between the second and third indices because it is manifest that
$A_{2,L}^{\;\;\;\;IJK}=-A_{2,L}^{\;\;\;\;IKJ}$,by definition.
Moreover there exists a symmetry between the upper indices:
$A_{2,L}^{\;\;\;\;IJK}=A_{2,L}^{\;\;\;\;JKI}=A_{2,L}^{\;\;\;\;KIJ}$.
Recall that in the supersymmetric transformation rules (\ref{susy}), 
$A_{2, L}^{\;\;\;\;\;IJK}$ appears in the second equation given by 
spin-$1/2$ fields. It was inevitable
to rewrite it in terms of superpotential in order to find out domain-wall 
solutions. Therefore we explicitly emphasize them here.

The kinetic terms (\ref{aijkl}) can be summarized as following seven 
$4\times 4$ block diagonal hermitian matrices like as 28-beins 
 $u^{IJ}_{\;\;\;KL}$ and $v^{IJKL}$ :
\bea 
A^{\;\;IJKL}_{\mu} = \mbox{diag}
(A_{\mu,1},A_{\mu,2},A_{\mu,3},A_{\mu,4},A_{\mu,5},A_{\mu,6},A_{\mu,7}),
\nonu 
\eea 
where each hermitian submatrix can be expressed as with empty space for
lower triangle parts
\bea 
\label{kinsu3}
& & A_{\mu,1} = \left(
\begin{array}{ccccc}
       & \left[12 \right] & \left[34 \right] & \left[56 \right] & \left[78
\right] \nonu \\  
 \left[12 \right]  & 0     &  -a^{\ast} & -a^{\ast} & -a    \nonu \\
\left[34 \right] & &     0   & -a^{\ast} & -a    \nonu \\
\left[56 \right] & &   & 0 & -a  \nonu \\
\left[78 \right] &   &  &  & 0 \end{array} \right),
 A_{\mu,2} =
\left(
\begin{array}{ccccc}
& \left[13 \right] & \left[24 \right] & \left[57 \right] & \left[68 \right] 
\nonu \\  
\left[13 \right] & 0     &  a^{\ast} & b^{\ast} & -b    \nonu \\
\left[24 \right] & &     0   & -b^{\ast} & b    \nonu \\
\left[57 \right] &   &   & 0 & a  \nonu \\
\left[68 \right] &  &  &  & 0 \end{array} \right), \nonu \\
& & A_{\mu,3} =  \left(
\begin{array}{ccccc}
& \left[14 \right] & \left[23 \right]  & \left[58 \right] & \left[67 \right] 
\nonu \\
\left[14 \right] &   0     &  -a^{\ast} & -b & -b^{\ast}    \nonu \\
\left[23 \right] & &     0   & -b & -b^{\ast}    \nonu \\
\left[58 \right] &   &   & 0 & -a \nonu \\
\left[67 \right] &  & &  & 0 \end{array} \right),
 A_{\mu,4} =
\left(
\begin{array}{ccccc}
 & \left[15 \right]  & \left[26 \right] & \left[37 \right] & \left[48 
\right] \nonu \\
\left[15 \right] &  0     &  a^{\ast} & -b^{\ast} & b    \nonu \\
\left[26 \right]  &  &     0   & b^{\ast} & -b    \nonu \\
\left[37 \right]  &   &   & 0 & a \nonu \\
\left[48 \right]  &   &  &  &  0 \end{array} \right), \nonu \\ 
& &
 A_{\mu,5} =
\left(
\begin{array}{ccccc}
 & \left[ 16 \right] & \left[25 \right] & \left[38 \right] & \left[47 
\right]  \nonu \\
\left[16 \right] &  0     &  -a^{\ast} & b & b^{\ast}    \nonu \\
\left[25 \right] & &     0   & b & b^{\ast}    \nonu \\
\left[38 \right] &   &   & 0 & -a  \nonu \\
\left[47 \right] &  &  &  & 0 \end{array} \right),
 A_{\mu,6} =
\left(
\begin{array}{ccccc}
 & \left[17 \right] & \left[28 \right] & \left[35 \right] & \left[46 
\right] \nonu \\
\left[17 \right] &   0     &  a & b^{\ast} & -b^{\ast}    \nonu \\
\left[28 \right] &   &     0   & -b & b    \nonu \\
\left[35 \right] &   &   & 0 & a^{\ast}\nonu \\
\left[46 \right] 
&  &  &  &  0 \end{array} \right), \nonu \\ 
& &
A_{\mu,7} =
\left(
\begin{array}{ccccc}
 & \left[18 \right] & \left[27 \right]  & \left[36 \right] & 
\left[45 \right] \nonu \\
\left[18 \right] &   0     &  -a & -b & -b    \nonu \\
 \left[27 \right] &   &  0   & -b^{\ast} & -b^{\ast}    \nonu \\
 \left[36 \right] &   &   & 0 & -a^{\ast} \nonu \\
\left[45 \right]  &  &  &   & 0  
\end{array} \right), \\
\eea
and the nonzero components are given by 
\bea 
a \equiv \frac{1}{2} e^{i\alpha} \left( \partial_{\mu}\lambda
 + i \sqrt{2} \; s \partial_{\mu} \al  \right), \qquad
 b \equiv 
\frac{1}{2} e^{i\phi} \left( \partial_{\mu} \lambda' + i \sqrt{2} s' \;
\partial_{\mu} \phi  \right). \nonu 
\eea

\section{Appendix B: $SO(5)$ Invariant Sectors}

The 28-beins $u^{IJ}_{\;\;\;KL}$ and $v^{IJKL}$ 
fields which are elements of $56 \times 56$ ${\cal V}(x)$ 
(\ref{56bein}) of
the fundamental 56-dimensional representation of $E_7$
 can be obtained 
by exponentiating the vacuum expectation values $\phi_{ijkl}$ 
(\ref{phiijklso5}) and (\ref{so5-}) 
of $SO(5)$ singlet space via (\ref{calV}) simultaneously. 
After tedious 
calculation  they have the
following seven $4 \times 4$ block diagonal hermitian matrices respectively:
\bea 
&&u_{1}^{\pm}
= \left(
\begin{array}{ccccc}
 & \left[12 \right] & \left[34 \right] & \left[56 \right] &
\left[78 \right] \nonu \\
 \left[12 \right] &  pq^2 & \pm p \be^2 & \pm \al q \be &  \pm
\al q \be \nonu \\
\left[34 \right]  &  & pq^2 & \al q \be & \al q \be \nonu \\
 \left[ 56\right] &  & & pq^2 &
p \be^2 \nonu \\
 \left[ 78\right] & &  &  & pq^2 \\
\end{array}
\right), 
u_{2}^{\pm} = \left(
\begin{array}{ccccc}
& \left[13 \right] & \left[24 \right] & \left[57 \right] &
\left[68 \right] \nonu \\
 \left[13 \right] & pr^2 & \mp p \ga^2 & \mp \al \ga r &  \pm 
\al \ga r \nonu \\
\left[24 \right]  &  & p r^2 & \al \ga r  & -\al \ga r  \nonu \\
 \left[ 57\right] &  &  & pr^2 & -p\ga^2 \nonu \\
 \left[ 68\right] &   &  & & p r^2 \\
\end{array}
\right), \nonu \\ 
&& u_{3}^{\pm} = \left(
\begin{array}{ccccc}
& \left[14 \right] & \left[23 \right] & \left[58 \right] &
\left[67 \right] \nonu \\
 \left[14 \right] & ps^2  & \pm p \de^2 & \mp \al s \de  & \mp 
\al s \de   \nonu \\
\left[23 \right]  &   &  p s^2 & -\al s \de & -\al s \de \nonu \\
 \left[ 58 \right] &   &  
& p s^2  & p \de^2 \nonu \\
 \left[ 67 \right] &   & 
&  &  ps^2 \\
\end{array}
\right), u_{4}^{\pm} = \left(
\begin{array}{ccccc}
& \left[15 \right] & \left[26 \right] & \left[37 \right] &
\left[48 \right] \nonu \\
 \left[15 \right] & qrs  & \pm q \ga \de & \mp \be r \de & \mp 
\be \ga s  \nonu \\
\left[26 \right]  &   & q r s  & -\be \ga s  & -\be r \de \nonu \\
 \left[37\right] &  &  & qrs  & q \ga \de 
\nonu \\
 \left[48 \right] &  &  &   & qrs  \\
\end{array}
\right), \nonu \\ 
&& u_{5}^{\pm} = \left(
\begin{array}{ccccc}
& \left[16 \right] & \left[25 \right] & \left[38 \right] &
\left[47 \right] \nonu \\
 \left[16 \right] & qrs  & \mp q\ga \de & \pm \be r \de & \mp 
\be \ga s  \nonu \\
\left[25 \right]  &  & qrs & -\be \ga s  & \be r \de \nonu \\
 \left[38 \right] &   &  & qrs  & -q \ga \de  \nonu \\
 \left[47 \right] &  &   
&  & qrs \\
\end{array}
\right), 
u_{6}^{\pm} = \left(
\begin{array}{ccccc}
& \left[17 \right] & \left[28 \right] & \left[35 \right] &
\left[46 \right] \nonu \\
 \left[17 \right] & qrs  & \pm q \ga \de & \pm \be r \de  & \pm
\be \ga s  \nonu \\
\left[28 \right]  &   & qrs & \be \ga s & \be r \de \nonu \\
 \left[35 \right] &  &  & qrs & q\ga \de \nonu \\
 \left[46 \right] &  &  &  & qrs  \\
\end{array}
\right), \nonu \\ 
&& u_{7}^{\pm} = \left(
\begin{array}{ccccc}
& \left[18 \right] & \left[27 \right] & \left[36 \right] &
\left[45 \right] \nonu \\
 \left[18 \right] & qrs  & \mp q \ga \de & \mp \be r \de & \pm
\be \ga s \nonu \\
\left[27 \right]  &  & qrs & \be \ga s & -\be r \de\nonu \\
 \left[36 \right] &  & & qrs  & -q \ga \de \nonu \\
 \left[45 \right] &  &  &  & qrs 
\end{array} \right), 
\nonu
\eea
\bea
 v_{1}^{\pm} &=& \varepsilon_{\pm}\left(
\begin{array}{ccccc}
& \left[12 \right] & \left[34 \right] & \left[56 \right] &
\left[78 \right] \nonu \\
 \left[12 \right] & \al \be^2 & \pm \al q^2 & \pm pq\be  
& \pm pq\be \nonu \\
\left[34 \right]  &  & \al\be^2 & pq\be
& pq\be \nonu \\
 \left[ 56\right] &  &  & \al \be^2 &
\al q^2 \nonu \\
 \left[ 78\right] &  &   &  & 
\al \be^2 \\
\end{array}
\right), \nonu \\ 
 v_{2}^{\pm} &=&  \varepsilon_{\pm}\left(
\begin{array}{ccccc}
& \left[13 \right] & \left[24 \right] & \left[57 \right] &
\left[68 \right] \nonu \\
 \left[13 \right] & \al \ga^2  &\mp \al r^2  & \mp p\ga r 
& \pm p r \ga \nonu \\
\left[24 \right]  &  & \al \ga^2 & pr \ga &
-p r \ga \nonu
\\
 \left[ 57\right] &   &  &  
\al \ga^2 &  -\al r^2 \nonu \\
 \left[ 68\right] &   & & & \al \ga^2  \\
\end{array}
\right), \nonu \\
 v_{3}^{\pm} &=& \varepsilon_{\pm} \left(
\begin{array}{ccccc}
& \left[14 \right] & \left[23 \right] & \left[58 \right] &
\left[67 \right] \nonu \\
 \left[14 \right] & \al \de^2 & \pm \al s^2 & \mp p \de s 
&  \mp p \de s \nonu \\
\left[23 \right]  &   &  \al \de^2 & -p\de s &
-p \de s \nonu \\
 \left[ 58 \right] &  &  & \al \de^2 &  \al s^2 \nonu \\
 \left[ 67 \right] &   &  &   & 
\al \de^2 \\
\end{array}
\right), \nonu \\ 
 v_{4}^{\pm} &=& \varepsilon_{\pm} \left(
\begin{array}{ccccc}
& \left[15 \right] & \left[26 \right] & \left[37 \right] &
\left[48 \right] \nonu \\
 \left[15 \right] & -\be \ga \de & \mp \be r s & \pm q \ga s & \pm 
q r \de  \nonu \\
\left[26 \right]  &   &  -\be \ga \de & q r \de 
& q \ga s \nonu
\\
 \left[37\right] &  &  & 
-\be \ga \de & -\be r s \nonu \\
 \left[48 \right] &  &  &  & -\be \ga \de \\
\end{array}
\right), \nonu \\
 v_{5}^{\pm} &=& \varepsilon_{\pm} \left(
\begin{array}{ccccc}
& \left[16 \right] & \left[25 \right] & \left[38 \right] &
\left[47 \right] \nonu \\
 \left[16 \right] & -\be \ga \de & \pm \be r s & \mp q \ga s  
& \pm q r \de   \nonu \\
\left[25 \right]  &  &-\be \ga \de  & q r \de  &  -q \ga s \nonu
\\
 \left[38 \right] &  &  & -\be \ga \de & \be r s  \nonu \\
 \left[47 \right] &  &  &  & -\be \ga \de 
 \\
\end{array}
\right), \nonu \\ 
 v_{6}^{\pm} &=& \varepsilon_{\pm} \left(
\begin{array}{ccccc}
& \left[17 \right] & \left[28 \right] & \left[35 \right] &
\left[46 \right] \nonu \\
 \left[17 \right] & -\be \ga \de & \mp \be r s & \mp q \ga s 
 & \mp q r \de  \nonu \\
\left[28 \right]  &  & -\be \ga \de & -q r \de
 & -q \ga s \nonu \\
 \left[35 \right] &  &  & -\be \ga \de  & -\be r s 
 \nonu \\
 \left[46 \right] &   &    &   & 
-\be \ga \de \\
\end{array}
\right), \nonu \\
 v_{7}^{\pm} &=& \varepsilon_{\pm} \left(
\begin{array}{ccccc}
& \left[18 \right] & \left[27 \right] & \left[36 \right] &
\left[45 \right] \nonu \\
 \left[18 \right] & -\be \ga \de  & \pm \be r s &  \pm q \ga s
 & \mp q r \de \nonu \\
\left[27 \right]  &  & -\be \ga \de  & -q r \de    
 &  q\ga s \nonu \\
 \left[36 \right] &  &   & -\be \ga \de 
& \be r s \nonu \\
 \left[45 \right] &   &  &      
& -\be \ga \de \\
\end{array} \right), \\
\label{uvso5+}
\eea 
where we denote the following combinations for simplicity
\bea
& & p \equiv \cosh\left(\frac{\la +\mu+ \rho}{2 \sqrt{2}} \right), \al \equiv
 \sinh\left(\frac{\la +\mu + \rho}{2 \sqrt{2}} \right),
q \equiv \cosh\left(\frac{\la}{2 \sqrt{2}} \right),
\be \equiv \sinh\left(\frac{\la }{2 \sqrt{2}} \right), \nonu \\
& & r \equiv \cosh\left(\frac{\mu}{2 \sqrt{2}} \right),
\ga \equiv \sinh\left(\frac{\mu }{2 \sqrt{2}} \right),
s \equiv \cosh\left(\frac{\rho}{2 \sqrt{2}} \right),
\de \equiv \sinh\left(\frac{\rho }{2 \sqrt{2}} \right), \nonu
\eea
and $ \varepsilon_{+}=1$ and $ \varepsilon_{-}=i$. We hope 
these quantities have no relations with the one in (\ref{pqrt}).

$\bullet$ $SO(5)^{+}$ embedding

They have the
following seven $4 \times 4$ block diagonal matrices respectively:
\bea 
u^{IJ}_{\;\;\;KL} & = & \mbox{diag}
(u_1^{+},u_2^{+},u_3^{+},u_4^{+},u_5^{+},u_6^{+},u_7^{+}), \nonu \\
v^{IJKL} & = & 
\mbox{diag} (v_1^{+},v_2^{+},v_3^{+}, v_4^{+}, v_5^{+}, v_6^{+}, v_7^{+}), 
\label{so5++}
\eea 
where the submatrices are in (\ref{uvso5+}).
The nonzero components of $A_2$ tensor, $A_{2, L}^{\;\;\;\;\;IJK}$, 
can be obtained from
(\ref{ttensor}) by simply plugging (\ref{uvso5+}) and  (\ref{so5++})
they are classified by four different fields with degeneracies 8, 16, 16,
16 respectively  and  
given by:
\bea
& & A_{2, 4}^{\;\;\;\;\;\;312}=A_{2, 2}^{\;\;\;\;\;134}= A_{2,
6}^{\;\;\;\;\;578}= A_{2, 1}^{\;\;\;\;\;\;324}=A_{2,
5}^{\;\;\;\;\;768}= A_{2, 3}^{\;\;\;\;\;214}= A_{2,
7}^{\;\;\;\;\;\;658}=A_{2, 8}^{\;\;\;\;\;567} \equiv y_{1,+},\nonu \\
& & A_{2, 5}^{\;\;\;\;\;612}= A_{2, 7}^{\;\;\;\;\;812}= A_{2,
5}^{\;\;\;\;\;\;634} =
A_{2, 7}^{\;\;\;\;\;834}= A_{2, 1}^{\;\;\;\;\;256}= A_{2,
3}^{\;\;\;\;\;\;456} =
A_{2, 4}^{\;\;\;\;\;356}= A_{2, 1}^{\;\;\;\;\;278}= \nonu \\
& & A_{2,
3}^{\;\;\;\;\;\;478} =
A_{2, 6}^{\;\;\;\;\;215}= A_{2, 8}^{\;\;\;\;\;437}= A_{2,
2}^{\;\;\;\;\;\;516} =
A_{2, 4}^{\;\;\;\;\;738}= A_{2, 8}^{\;\;\;\;\;217}= A_{2,
6}^{\;\;\;\;\;\;435}=A_{2, 2}^{\;\;\;\;\;718} \equiv y_{2,+}, \nonu \\
& & A_{2, 5}^{\;\;\;\;\;814}= A_{2, 6}^{\;\;\;\;\;714}= A_{2,
5}^{\;\;\;\;\;\;823} =
A_{2, 6}^{\;\;\;\;\;723}= A_{2, 1}^{\;\;\;\;\;458}= A_{2,
2}^{\;\;\;\;\;\;358} =
A_{2, 1}^{\;\;\;\;\;467}= A_{2, 2}^{\;\;\;\;\;367}= \nonu \\
& & A_{2,
8}^{\;\;\;\;\;\;415} =
A_{2, 7}^{\;\;\;\;\;326}= A_{2, 7}^{\;\;\;\;\;416}= A_{2,
8}^{\;\;\;\;\;\;325} =
A_{2, 4}^{\;\;\;\;\;617}= A_{2, 3}^{\;\;\;\;\;528}= A_{2,
4}^{\;\;\;\;\;\;518}=A_{2, 3}^{\;\;\;\;\;627} \equiv y_{3,+}, \nonu \\
& & A_{2, 5}^{\;\;\;\;\;713}= A_{2, 8}^{\;\;\;\;\;613}= A_{2,
6}^{\;\;\;\;\;\;824} =
A_{2, 7}^{\;\;\;\;\;524}= A_{2, 5}^{\;\;\;\;\;427}= A_{2,
1}^{\;\;\;\;\;\;757} =
A_{2, 4}^{\;\;\;\;\;257}= A_{2, 2}^{\;\;\;\;\;468}= \nonu \\
& & A_{2,
7}^{\;\;\;\;\;\;315} =
A_{2, 8}^{\;\;\;\;\;426}= A_{2, 3}^{\;\;\;\;\;816}= A_{2,
1}^{\;\;\;\;\;\;638} =
A_{2, 2}^{\;\;\;\;\;547}= A_{2, 3}^{\;\;\;\;\;517}= A_{2,
4}^{\;\;\;\;\;\;628}=A_{2, 6}^{\;\;\;\;\;318} \equiv y_{4,+}. \nonu 
\eea
Here they have simple form:
\bea
y_{1,+} 
&  = &  \frac{1}{8\sqrt{uvw}} \left(-3 + u^2v^2 + 
u^2w^2+v^2w^2
 \right),  
\nonu \\
y_{2,+} & = &\frac{1}{8\sqrt{uvw}} \left(1 - u^2v^2 - 
u^2w^2+v^2w^2
 \right),   
%
\nonu \\
y_{3,+} & = &  \frac{1}{8\sqrt{uvw}} \left(-1 - u^2v^2 + 
u^2w^2+v^2w^2
 \right),   
\nonu \\
y_{4,+} & = &
\frac{1}{8\sqrt{uvw}} \left(-1 + u^2v^2 -
u^2w^2+ v^2w^2
 \right).    
\label{so5y+}
\eea
As before in $SU(3)$ invariant sectors, 
there exists a symmetry between the upper indices:
$A_{2,L}^{\;\;\;\;IJK}=A_{2,L}^{\;\;\;\;JKI}=A_{2,L}^{\;\;\;\;KIJ}$.
The kinetic terms can be summarized as following block diagonal hermitian
matrices:
\bea
& & A_{\mu,1} = \left(
\begin{array}{ccccc}
       & \left[12 \right] & \left[34 \right] & \left[56 \right] & \left[78
\right] \nonu \\  
 \left[12 \right]  & 0     &  -K & -A & -A    \nonu \\
\left[34 \right] &  &     0   & -A & -A    \nonu \\
\left[56 \right] &   &   & 0 & -K  \nonu \\
\left[78 \right] &   &  & & 0 \end{array} \right),
 A_{\mu,2} =
\left(
\begin{array}{ccccc}
& \left[13 \right] & \left[24 \right] & \left[57 \right] & \left[68 \right] 
\nonu \\  
\left[13 \right] & 0     & K  & B  & -B    \nonu \\
\left[24 \right] &  &     0   & -B & B    \nonu \\
\left[57 \right] &    &   & 0 & K  \nonu \\
\left[68 \right] &  &  &  & 0 \end{array} \right),  \\
& & A_{\mu,3} =  \left(
\begin{array}{ccccc}
& \left[14 \right] & \left[23 \right]  & \left[58 \right] & \left[67 \right] 
\nonu \\
\left[14 \right] &   0     &  -K & F & F    \nonu \\
\left[23 \right] &  &     0   & F & F    \nonu \\
\left[58 \right] &   &   & 0 & -K \nonu \\
\left[67 \right] &  &  &  & 0 \end{array} \right),
 A_{\mu,4} =
\left(
\begin{array}{ccccc}
 & \left[15 \right]  & \left[26 \right] & \left[37 \right] & \left[48 
\right] \nonu \\
\left[15 \right] &  0     &  A & -B & -F    \nonu \\
\left[26 \right]  &  &     0   & -F & -B    \nonu \\
\left[37 \right]  &   &  & 0 & A \nonu \\
\left[48 \right]  &   &  &  &  0 \end{array} \right),  \\ 
& &
A_{\mu,5} =
\left(
\begin{array}{ccccc}
 & \left[ 16 \right] & \left[25 \right] & \left[38 \right] & \left[47 
\right]  \nonu \\
\left[16 \right] &  0     &  -A & B & -F    \nonu \\
\left[25 \right] & &     0   & -F & B    \nonu \\
\left[38 \right] &   &   & 0 & -A  \nonu \\
\left[47 \right] &  &  &  & 0 \end{array} \right),
 A_{\mu,6} =
\left(
\begin{array}{ccccc}
 & \left[17 \right] & \left[28 \right] & \left[35 \right] & \left[46 
\right] \nonu \\
\left[17 \right] &   0     &  A & B & F    \nonu \\
\left[28 \right] &   &     0   & F & B    \nonu \\
\left[35 \right] &   &   & 0 & A\nonu \\
\left[46 \right] 
&  &  &  &  0 \end{array} \right),  \\ 
& &
A_{\mu,7} =
\left(
\begin{array}{ccccc}
 & \left[18 \right] & \left[27 \right]  & \left[36 \right] & 
\left[45 \right]  \\
\left[18 \right] &   0     &  -A & -B & F     \\
 \left[27 \right] &   &  0   & F & -B     \\
 \left[36 \right] &   &   & 0 & -A  \\
\left[45 \right]  &  &  &   & 0 
\end{array} \right), 
\label{kineticso5}
\eea
and nonzero components are 
\bea 
A \equiv \frac{1}{2} \partial_{\mu}\lambda, \qquad
B \equiv \frac{1}{2}  \partial_{\mu} \mu, \qquad 
F \equiv \frac{1}{2} \pa_{\mu} \rho, \qquad
K \equiv \frac{1}{4} \left(\partial_{\mu}\lambda + \partial_{\mu} \mu +
 \pa_{\mu} \rho \right). \nonu 
\eea

$\bullet$ $SO(5)^{-}$ embedding

The 28-beins $u^{IJ}_{\;\;\;KL}$ and $v^{IJKL}$ 
fields which are elements of $56 \times 56$ ${\cal V}(x)$ 
(\ref{56bein}) of
the fundamental 56-dimensional representation of $E_7$
 can be obtained 
by exponentiating the vacuum expectation values $\phi_{ijkl}$ 
(\ref{so5-}) of $SO(5)$ singlet space via (\ref{calV}). After tedious 
calculation  they have the
following seven $4 \times 4$ block diagonal hermitian matrices respectively:
\bea 
u^{IJ}_{\;\;\;KL} & = & \mbox{diag}
(u_1^{-},u_2^{-},u_3^{-},u_4^{-},u_5^{-},u_6^{-},u_7^{-}), \nonu \\
v^{IJKL} & = & 
\mbox{diag} (v_1^{-},v_2^{-},v_3^{-}, v_4^{-}, v_5^{-}, v_6^{-}, v_7^{-}), 
\label{uvso5-} 
\eea 
where the submatrices
are the same as those in (\ref{uvso5+}).
In this case, $A^{\;\;IJKL}_{\mu}$ is $-i$ times the one of $SO(5)^{+}$ 
invariant case.
Moreover, 
The nonzero components of $A_2$ tensor, $A_{2, L}^{\;\;\;\;\;IJK}$ 
can be obtained from
(\ref{ttensor}) by simply plugging (\ref{uvso5-}) and (\ref{uvso5+}). 
They are classified by four different fields with degeneracies 8, 16, 16,
16 respectively  and  
given by
\bea
y_{1,-} & = & \frac{(1+i)}{16(uvw)^{3/2}} \left(
 -i u^2 + u^3 v^3 w +\mbox{two cyclic permutations}-3 uvw+3 i u^2v^2w^2  
\right), \nonu \\
y_{2,-} & = &  \frac{(1+i)}{16(uvw)^{3/2}} \left(
 -i u^2+i v^2 +i w^2  - u^3 v^3 w -u^3vw^3 +u v^3w^3  + uvw- i u^2v^2w^2  
\right), \nonu \\
y_{3,-} & = & 
 \frac{(1+i)}{16(uvw)^{3/2}} \left(
 i u^2-i v^2 +i w^2  - u^3 v^3 w + u^3vw^3 - u v^3w^3  + uvw- i u^2v^2w^2  
\right), \nonu \\
y_{4,-} & = &  \frac{(1+i)}{16(uvw)^{3/2}} \left(
 -i u^2-i v^2 +i w^2  - u^3 v^3 w + u^3vw^3 + u v^3w^3  - uvw+ i u^2v^2w^2  
\right). 
\label{so5y-} 
\eea

\section{Appendix C: $SO(3) \times SO(3)$ Invariant Sector}

The 28-beins $u^{IJ}_{\;\;\;KL}$ and $v^{IJKL}$ 
fields which are elements of $56 \times 56$ ${\cal V}(x)$ 
(\ref{56bein}) of
the fundamental 56-dimensional representation of $E_7$
 can be obtained 
by exponentiating the vacuum expectation values $\phi_{ijkl}$ 
(\ref{so3so3ijkl}) of $SO(3) \times SO(3)$ 
singlet space via (\ref{calV}). After tedious 
calculation  they have the
following thirteen $4 \times 4$ matrices respectively:
\bea 
u^{IJ}_{\;\;\;KL}  = 
\left(
\begin{array}{ccccccc}
u_1 & 0  & 0 & 0 & 0 & u_2 & 0 \nonu \\
    0 & u_3  & 0 & 0 & u_4 & 0 & 0 \nonu \\
0 & 0  & u_5 & u_6 & 0 & 0 & 0 \nonu \\
0 & 0  & u_7 & u_8 & 0 & 0 & 0 \nonu \\
0 & u_9  & 0 & 0 & u_{10} & 0 & 0 \nonu \\
u_{11} & 0  & 0 & 0 & 0 & u_{12} & 0 \nonu \\
0 & 0  & 0 & 0 & 0 & 0 & u_{13}
\end{array}
\right), \nonu
\eea
\bea
v^{IJKL}  = 
\left(
\begin{array}{ccccccc}
v_1 & 0  & 0 & 0 & 0 & v_2 & 0 \nonu \\
    0 & v_3  & 0 & 0 & v_4 & 0 & 0 \nonu \\
0 & 0  & v_5 & v_6 & 0 & 0 & 0 \nonu \\
0 & 0  & v_7 & v_8 & 0 & 0 & 0 \nonu \\
0 & v_9  & 0 & 0 & v_{10} & 0 & 0 \nonu \\
v_{11} & 0  & 0 & 0 & 0 & v_{12} & 0 \nonu \\
0 & 0  & 0 & 0 & 0 & 0 & v_{13} \\
\end{array}
\right).
\nonu 
\eea 
Each submatrix is $4 \times 4$ matrix and we denote
antisymmetric index pairs. Since they are very complicated expressions in terms 
of hyperbolic functions of $\la^{\al}$, we do not list them here. 
For explicit forms we refer to the original version in the hep-th archive.

The kinetic terms (\ref{aijkl}) restricted to the 
scalar manifold satisfying the constraint $\la^{\al}=0, \al=2,\cdots,7 $
can be summarized as following nine 
$4\times 4$ matrices:
\bea
A^{\;\;IJKL}_{\mu}  = 
\left(
\begin{array}{ccccccc}
A_{\mu,1} & 0  & 0 & 0 & 0 & A_{\mu,2} & 0  \\
    0 & A_{\mu,3}  & 0 & 0 & A_{\mu,4} & 0 & 0  \\
0 & 0  & A_{\mu,5} & A_{\mu,6} & 0 & 0 & 0  \\
0 & 0  & 0 & A_{\mu,8} & 0 & 0 & 0 \\
0 & 0  & 0 & 0 & A_{\mu,10} & 0 & 0  \\
0 & 0  & 0 & 0 & 0 & A_{\mu,12} & 0  \\
0 & 0  & 0 & 0 & 0 & 0 & 0 
\end{array}
\right),
\label{so3so3kinetic}
\eea
where each submatrix has the following forms:
\bea 
&& A_{\mu,1}
= \left(
\begin{array}{ccccc}
 & \left[12 \right] & \left[34 \right] & \left[56 \right] &
\left[78 \right] \nonu \\
 \left[12 \right] & 0  & -a  & 0
 & 0 \nonu \\
\left[34 \right]  &-a  & 0 & 0 &0  \nonu \\
 \left[ 56\right] & 0 & 0  & 0  &-a
 \nonu \\
 \left[ 78\right] & 0 & 0  &- a & 0  \\
\end{array}
\right), A_{\mu,2} = \left(
\begin{array}{ccccc}
& \left[17 \right] & \left[28 \right] & \left[35 \right] &
\left[46 \right] \nonu \\
 \left[12 \right]
& 0 &0  & -i a
 & 0 \nonu \\
\left[34 \right]  & 0 & 0 & 0 & 0 \nonu \\
 \left[ 56\right] & 0 & 0  & 0  &
0 \nonu \\
 \left[ 78\right] &0 & 0  & 0 & -i a   \\
\end{array}
\right), \nonu \\ 
&& A_{\mu,3} = \left(
\begin{array}{ccccc}
& \left[13 \right] & \left[24 \right] & \left[57 \right] &
\left[68 \right] \nonu \\
 \left[13 \right] & 0 & a & 0
 & 0 \nonu \\
\left[24 \right]  & a & 0 & 0 & 0 \nonu \\
 \left[ 57\right] & 0 & 0  & 0  & a
 \nonu \\
 \left[ 68\right] & 0 & 0  &a  & 0  \\
\end{array}
\right), A_{\mu,4} = \left(
\begin{array}{ccccc}
& \left[16 \right] & \left[25 \right] & \left[38 \right] &
\left[47 \right] \nonu \\
 \left[13 \right] & 0 & i a &0 
 & 0 \nonu \\
\left[24 \right]  & 0 & 0 & 0 & 0 \nonu \\
 \left[ 57\right] & 0 & 0  &0   &0
 \nonu \\
 \left[ 68\right] &0 & 0  & 0 & i a \\
\end{array}
\right), \nonu \\ 
&& A_{\mu,5} = \left(
\begin{array}{ccccc}
& \left[14 \right] & \left[23 \right] & \left[58 \right] &
\left[67 \right] \nonu \\
 \left[14 \right]  &0  & -a & 0
 &0  \nonu \\
\left[23 \right]  &-a  & 0 & 0 & 0 \nonu \\
 \left[ 58\right] & 0 &  0 & 0  &
-a \nonu \\
 \left[ 67\right] &0 & 0  & -a & 0  \\
\end{array}
\right), A_{\mu,6} = \left(
\begin{array}{ccccc}
& \left[15 \right] & \left[26 \right] & \left[37 \right] &
\left[48 \right] \nonu \\
 \left[14 \right] & 0  & 0 & 0
 & 0 \nonu \\
\left[23 \right]  & -i a & 0 & 0 & 0 \nonu \\
 \left[ 58\right] &0  & 0  & 0  &0
 \nonu \\
 \left[ 67\right] &0 & 0  & 0 & -i a  \\
\end{array}
\right), \nonu \\ 
&& A_{\mu,8} = \left(
\begin{array}{ccccc}
& \left[14 \right] & \left[23 \right] & \left[58 \right] &
\left[67 \right] \nonu \\
 \left[15 \right]  & 0 & -i a & 0  
 &0  \nonu \\
\left[26 \right]  & 0 &0  & 0 & 0 \nonu \\
 \left[ 37\right] & 0  & 0  & 0  &0
 \nonu \\
 \left[ 48\right] &0 & 0  &0  & -i a  \nonu \\
\end{array}
\right), 
A_{\mu,10} = \left(
\begin{array}{ccccc}
& \left[13 \right] & \left[24 \right] & \left[57 \right] &
\left[68 \right] \nonu \\
 \left[16 \right] & 0 &0  &0 
 & 0 \nonu \\
\left[25 \right]  & i a & 0 & 0 & 0 \nonu \\
 \left[ 38\right] & 0 & 0  & 0  &0
 \nonu \\
 \left[ 47\right] &0 & 0  & 0 & i a  \\
\end{array}
\right), \nonu \\  
& & A_{\mu,12} = \left(
\begin{array}{ccccc}
& \left[12 \right] & \left[34 \right] & \left[56 \right] &
\left[78 \right] \nonu \\
 \left[17 \right]  & 0 & 0 &0 
 & 0 \nonu \\
\left[28 \right]  & 0 & 0 & 0 & 0 \nonu \\
 \left[ 35\right] &-i a  & 0  &0   &0
 \nonu \\
 \left[ 46\right] &0 & 0  &0  &-i a   \\
\end{array}
\right),    
\eea
where nonzero component is 
\bea
a \equiv \frac{1}{4} ( 1+\cosh \la_1 ) \pa_{\mu} \la_1.
\nonu
\eea

\vspace{2cm}
\centerline{\bf Acknowledgments} 

This research was supported 
by Korean Physical Society(1999),
Kyungpook National University Research Fund, 2000 and
grant 2000-1-11200-001-3 from the Basic Research Program of the Korea
Science $\&$ Engineering Foundation.


\begin{thebibliography}{[00]}
\bibitem{ar} C. Ahn and S.-J. Rey, 
Nucl.Phys. {\bf B565} (2000) 210.
\bibitem{ar1} C. Ahn and S.-J. Rey, 
Nucl.Phys. {\bf B572} (2000) 188.
\bibitem{maldacena} J. Maldacena, Adv.Theor.Math.Phys. {\bf 2} (1998) 231.
\bibitem{witten} E. Witten, Adv.Theor.Math.Phys. {\bf 2} (1998) 253.
\bibitem{gkp} S.S. Gubser, I.R. Klebanov and A.M. Polyakov,
Phys.Lett. {\bf 428B} (1998) 105.
\bibitem{duff} M.J. Duff, B.E.W. Nilsson and C.N. Pope, Phys. Rep. 
{\bf 130} (1986) 1.
\bibitem{dnw} B. de Wit, H. Nicolai and N.P. Warner, Nucl.Phys. {\bf B255}
(1985) 29.
\bibitem{dp} M.J. Duff and C.N. Pope, {\it Kaluza-Klein supergravity and
the seven sphere}, in: Supersymmetry and Supergravity '82, eds. S. Ferrara,
J.G. Taylor and P. van Nieuwenhuizen(World Scientific, Singapore, 1983).
\bibitem{englert} F. Englert, Phys.Lett. {\bf B119} (1982) 339.
\bibitem{dn} B. de Wit and H. Nicolai, Phys.Lett. {\bf B148} (1984) 60.
\bibitem{pw} C.N. Pope and N.P. Warner, Phys.Lett. {\bf B150} (1985) 352.
\bibitem{dn1} B. de Wit and H. Nicolai, Nucl.Phys. {\bf B231} (1984) 506.
\bibitem{ap} C. Ahn and J. Paeng, 
{\tt hep-th/0008065}, Nucl.Phys. {\bf  B595} (2001) 119.
\bibitem{ar2} C. Ahn and S.-J. Rey, to appear.
\bibitem{warner} N.P. Warner, Phys.Lett. {\bf B128} (1983) 169.

\bibitem{dn3} B. de Wit and H. Nicolai, Phys.Lett. {\bf B108} (1982) 285.
\bibitem{dn2} B. de Wit and H. Nicolai, Nucl.Phys. {\bf B208} (1982) 323.
\bibitem{cremmer} E. Cremmer, B. Julia and J. Scherk, Phys.Lett. {\bf B76}
(1978) 409; E. Cremmer and B. Julia, Nucl.Phys. {\bf B159} (1979) 141.
\bibitem{cj} E. Cremmer and B. Julia, Phys.Lett. {\bf B80} (1978) 48.
\bibitem{warner1} N.P. Warner,
Nucl.Phys. {\bf B231} (1984) 250.
\bibitem{awada} M.A. Awada, M.J. Duff and C.N. Pope, Phys.Rev.Lett. {\bf 50}
(1983) 294; M.J. Duff, B.E.W. Nilsson and C.N. Pope, Phys.Rev.Lett. {\bf 50}
(1983) 2043; {\bf 51} (1983) 846(errata).
\bibitem{romans} L.J. Romans, Phys.Lett. {\bf B131} (1983) 83.
\bibitem{dn6} B. de Wit and H. Nicolai, Nucl.Phys. {\bf B188} (1981) 98.
\bibitem{clp}M. Cvetic, H. Lu and C.N. Pope,  Nucl.Phys. {\bf B574} (2000) 761.
\bibitem{cveticetal} M. Cvetic, S. Griffies and S.-J. Rey, Nucl.Phys.
{\bf B381} (1992) 301;
M. Cvetic, S. Griffies and S.-J. Rey, Nucl.Phys. {\bf B389} (1993) 3.
\bibitem{pilchwarner} K. Pilch and N.P. Warner, hep-th/0006066.
\bibitem{bf} P. Breitenlohner and D.Z. Freedman, Phys.Lett. {\bf B115}
(1982) 197; Ann.Phys. {\bf 144} (1982) 249.
\bibitem{st} K. Skenderis and P.K. Townsend, Phys.Lett. {\bf B468} (1999) 46.
\bibitem{dn86}  B. de Wit and H. Nicolai, Nucl.Phys. {\bf B281} (1987) 211.
\bibitem{domainwall} L. Girardello, M. Petrini, M. Porrati and A. Zaffaroni,
JHEP {\bf 9812} (1998) 022;
D.Z. Freedman, S.S. Gubser, K. Pilch and N.P. Warner, {\tt hep-th/9904017}.
\bibitem{distler} J. Distler and F. Zamora, Adv.Theor.Math.Phys. {\bf 2}
(1998) 1405.
\bibitem{boucher} W. Boucher, Nucl.Phys. {\bf B242} (1984) 282. 

\end{thebibliography}
\end{document}